\documentclass[aps,pre,onecolumn, superscriptaddress,nofootinbib]{revtex4-2}

\usepackage{graphicx}
\usepackage{amsmath}
\usepackage{amssymb}
\usepackage{color}
\usepackage{mathrsfs}
\usepackage[colorlinks,linkcolor=blue,urlcolor=cyan,citecolor=blue,anchorcolor=blue]{hyperref}
\usepackage[T1]{fontenc}

\usepackage{orcidlink}

\begin{document}

\title{Fractional Langevin equation far from equilibrium: Riemann-Liouville
fractional Brownian motion, spurious nonergodicity and aging}

\author{Qing Wei\,\orcidlink{0000-0001-5287-9430}} \email{weiqingbambi@gmail.com}
\affiliation{LSEC, ICMSEC, Academy of Mathematics and Systems Science, Chinese
Academy of Sciences, Beijing 100190, China}
\author{Wei Wang\,\orcidlink{0000-0002-1786-3932}}
\email{weiwangnuaa@gmail.com}
\affiliation{University of Potsdam, Institute of Physics \& Astronomy, 14476
Potsdam-Golm, Germany}
\author{Yifa Tang\,\orcidlink{0000-0002-3301-3487}}
\email{tyf@lsec.cc.ac.cn}
\affiliation{LSEC, ICMSEC, Academy of Mathematics and Systems Science, Chinese
Academy of Sciences, Beijing 100190, China}
\affiliation{School of Mathematical Sciences, University of Chinese Academy of
Sciences, Beijing 100049, China}
\author{Ralf Metzler\,\orcidlink{0000-0002-6013-7020}}
\email{rmetzler@uni-potsdam.de}
\affiliation{University of Potsdam, Institute of Physics \& Astronomy, 14476
Potsdam-Golm, Germany}
\affiliation{Asia Pacific Centre for Theoretical Physics, Pohang 37673,
Republic of Korea}
\author{Aleksei Chechkin\,\orcidlink{0000-0002-3803-1174}}
\email{chechkin@uni-potsdam.de}
\affiliation{University of Potsdam, Institute of Physics \& Astronomy, 14476
Potsdam-Golm, Germany}
\affiliation{Asia Pacific Centre for Theoretical Physics, Pohang 37673,
Republic of Korea}
\affiliation{Faculty of Pure and Applied Mathematics, Hugo Steinhaus Center,
Wroc{\l}aw University of Science and Technology, Wyspianskiego 27, 50-370
Wroc{\l}aw, Poland}
\affiliation{German-Ukrainian Core of Excellence, Max Planck Institute of
Microstructure Physics, Weinberg 2, 06120 Halle, Germany}

\begin{abstract}
We consider the fractional Langevin equation far from equilibrium (FLEFE) to
describe stochastic dynamics which do not obey the fluctuation-dissipation
theorem, unlike the conventional  fractional Langevin equation
(FLE). The solution of this equation is Riemann-Liouville fractional
Brownian motion (RL-FBM), also known in the literature as FBM II. Spurious
nonergodicity, stationarity, and aging properties of the solution are
explored for all admissible values $\alpha>1/2$ of the order $\alpha$ of
the time-fractional Caputo derivative in the FLEFE. The increments of the
process are asymptotically stationary. However when $1/2<\alpha<3/2$, the
time-averaged mean-squared displacement (TAMSD) does not converge to the
mean-squared displacement (MSD). Instead, it converges to the mean-squared
increment (MSI) or structure function, leading to the phenomenon of spurious
nonergodicity. When $\alpha\ge 3/2$, the increments of FLEFE motion
are nonergodic, however the higher order increments are asymptotically
ergodic. We also discuss the aging effect in the FLEFE by investigating
the influence of an aging time $t_a$ on the mean-squared displacement,
time-averaged mean-squared displacement and autocovariance function of the
increments. We find that under strong aging conditions the process becomes
ergodic, and the increments become stationary in the domain $1/2<\alpha<3/2$.
\end{abstract}

\maketitle

\section{Introduction}

In his famous note "On the Theory of Brownian Motion" of 1908 Paul Langevin
formulated Newton's second law for a test Brownian particle immersed in a
fluid or gas at equilibrium \cite{langevin1908}. In modern terms, Langevin's
stochastic differential equation for the position $x$ of the particle at time
$t$ is given by (in this paper we consider the one-dimensional case)
\cite{coffey2004,zwanzig2001, kampen1981}
\begin{equation}\label{SDE}
m\frac{d^2x(t)}{dt^2}=-m\eta\frac{dx(t)}{dt}+f(t),
\end{equation}
where $m$ is the mass of the particle and $\eta$ is friction coefficient with
dimension time$^{-1}$. The first term on the right-hand side represents the
frictional force exerted by the medium and the second term is the random
force $f(t)$ due to the random collisions of the surrounding molecules
with the test particle. In the theory of Brownian motion, the random
force is chosen as zero-mean white Gaussian noise with autocovariance
function (ACF) $\left\langle f(t)f\left(t^{\prime}\right)\right\rangle=2
\mathcal{K}\delta\left(t-t^{\prime}\right)$, where $\mathcal{K}$
is the noise intensity and $\delta(\cdot)$ is the Dirac delta function,
following $\int_{-\infty}^{\infty}\delta(\tau)d\tau = 1$. Importantly,
the friction and random forces are not independent: the noise intensity
and friction coefficient are related by $\mathcal{K} = k_BTm\eta$,
in which $k_B$ is the Boltzmann constant and $T$ is the temperature of the
gas or fluid in which the Brownian particle is immersed. The last relation
is the simplest example of the fluctuation-dissipation theorem (FDT) for a
particle in an equilibrated bath or at thermal equilibrium \cite{kubo1966}. A
noise obeying the FDT is called internal \cite{klimontovich1995} and otherwise
external. The overdamped form of the Langevin equation neglects the inertial
term on the left hand side of Eq.~(\ref{SDE}) and is used for the description
of Brownian motion in a medium with strong friction, that is, large viscosity,
\begin{equation}\label{LE1}
\frac{dx(t)}{d t} =\sqrt{2K}\xi(t).
\end{equation}
A typical example for a Brownian particle are micron-sized colloidal spheres
in water. In Eq.~(\ref{LE1}), $\xi(t)$ denotes zero-mean white Gaussian noise
with $\left<\xi(t)\xi(t^\prime)\right>=\delta(t-t^\prime)$ and $K$ of physical
dimension $\mathrm{length}^2/\mathrm{time}$ is the diffusion coefficient that
follows the Einstein relation or FDT $K=k_BT/(m\eta)$. Eq.~(\ref{LE1}) encodes
the familiar linear time dependence of the mean-squared displacement (MSD),
$\left<x^2(t)\right>=2Kt$, which is called normal diffusion. In what follows we
will deal with generalizations of the overdamped Langevin equation~(\ref{LE1}).

Over the past decades, anomalous diffusion phenomena characterized by a
non-linear dependence of the MSD have been found ubiquitously in nature
and studied intensively. Here we refer the reader to several  monographs
\cite{klages2008, pekalski1999, evangelista2023, sikorskii2019} as well
as reviews \cite{bouchaud1990, metzler2000, sokolov2012, hoefling2013,
metzler2014} and numerous references therein. In the most typical form
of anomalous diffusion with a power-law form $\langle x^2(t)\rangle\simeq
t^{\alpha}$ and scaling exponent $\alpha$. One distinguishes between slow
diffusion, or subdiffusion if the MSD grows sublinearly in time ($0<\alpha<1$),
and fast diffusion, or superdiffusion characterized by superlinear increase
of the MSD ($\alpha>1$). There are two generic Langevin-like models based
on generalizations of Eqs.~(\ref{SDE}) and (\ref{LE1}), which account for
anomalous diffusion in a lot of practical applications.

The first generalization is related to the fractional Brownian motion (FBM),
introduced by Kolmogorov \cite{kolmogorov1940}, and Mandelbrot and van Ness
\cite{mandelbrot1968}, which can be defined via the Langevin equation
\begin{equation}\label{MNFBM}
\frac{d}{d t} x(t)= \sqrt{2K_\alpha}\xi_\alpha(t),
\end{equation}
where $K_\alpha$ is the generalized diffusion coefficient with
physical dimension $\mathrm{length}^2/\mathrm{time}^{\alpha}$, and
$\xi_\alpha(t)$ is the fractional Gaussian noise, that is a stationary
Gaussian process with zero-mean and long-time power-law decay of
the ACF ($t^\prime\gg t$), $\left\langle\xi_\alpha\left(t\right)
\xi_\alpha\left(t^\prime\right)\right\rangle \sim \left[\alpha(\alpha-
1)/2\right] |t_1-t_2|^{\alpha-2}$ for which the anomalous diffusion exponent
$\alpha\in (0, 2]$, $\alpha \neq 1$. In the mathematical literature,
FBM is defined with the Hurst exponent $H=\alpha/2$. The ordinary BM in
Eq.~(\ref{LE1}), corresponds to the case $\alpha=1$ \cite{qian2003}. The
MSD of FBM grows like $\simeq t^\alpha$, so FBM accounts for both sub-
and superdiffusion phenomena \footnote{Note that in order to describe
subdiffusion, the integral from zero to infinity over the ACF of the noise
must be zero \cite{mandelbrot1968}.}. Importantly, the FBM (\ref{MNFBM}) does
not fulfill the FDT and thus cannot describe diffusion in systems close to
equilibrium. Instead, the noise is considered to be external which is immanent
to open systems \cite{klimontovich1995}, e.g., living cells \cite{weber2010,
jeon2011}, crowded fluids \cite{weiss2013, weiss2009}, movement ecology
\cite{vilk2022}, serotonergic brain fibers \cite{janusonis2023, janusonis2020},
or financial markets \cite{comte1998, marinucci1999, gatheral2018}.

The second model, the generalized Langevin equation (GLE) suggested by Mori
and Kubo \cite{mori1965, kubo1966, zwanzig2001}, provides a stochastic
description of thermalized systems near equilibrium. The overdamped form
reads \cite{mori1965, kubo1966}
\begin{equation}\label{gle}
\int_0^t \gamma(t-t^{\prime})\frac{{d} x\left(t^{\prime}\right)}{{d} t^{\prime}} d t^{\prime}=\zeta(t),
\end{equation}
in which $\gamma(t)$ is the friction kernel of dimension
time$^{-2}$ and $\zeta(t)$ is a Gaussian fluctuating driving
force whose ACF is  coupled to the friction kernel by the FDT
$\langle\zeta(t)\zeta(t^\prime)\rangle=[k_{B}T/m]\gamma(|t-t^\prime|)$. The
GLE reduces to the normal Langevin equation when $\gamma(t)=2\eta\delta(t)$
\cite{panja2010, kou2008}. In the cytoplasm of a biological cell or cell
extract, a particle moves through a medium characterized by macromolecular
crowding and the presence of elastic elements, which provides the cytoplasm
with viscoelastic properties. In other words, the cytoplasm "pushes
back" and ensures long-time correlations in the particle's trajectory
\cite{weber2010}. The particle then exhibits subdiffusive behavior,
which is modeled by the fractional Langevin equation (FLE) \cite{lutz2001}
that is a particular case of the GLE in Eq.~(\ref{gle}) with a power-law
friction kernel $\gamma_{\alpha} (t) \propto  t^{-\alpha}$, where $0<\alpha<1$
\cite{deng2009, molina2018}.

The FLE can be conveniently written in terms of the Caputo fractional
derivative of order $n-1<\alpha<n, n \in \mathbb{N}^{+}$, which is defined
as \cite{caputo1966, caputo1967}
\begin{equation} \label{caputo-derivative}
^{C}_{0}D^{\alpha}_{t}[f(t)] \equiv \frac{d^\alpha x(t)}{dt^\alpha} = \int_{0}^{t}f^{(n)}(u)\frac{(t-u)^{-\alpha+n-1}}{\Gamma(n-\alpha)}du.
\end{equation} 
In particular, when $\alpha = n\in \mathbb{N}_{0}$, the Caputo derivative is
reduced to the normal derivative $^{C}_{0}D^{n}_{t}[f(t)]=f^{(n)}(t)$
\cite{podlubny1999}.

Then the overdamped form of the FLE reads
\begin{equation}\label{fle}
\eta_\alpha \frac{d^\alpha x(t)}{dt^\alpha} = \zeta_\alpha(t), ~0<\alpha<1 ,
\end{equation}
where the noise $\zeta_{\alpha}(t)$ satisfies the FDT $\langle\zeta_{
\alpha}(t)\zeta_{\alpha}(t')\rangle=[k_BT\eta_{\alpha}/(m\Gamma(1-\alpha))]
|t-t'|^{-\alpha}$ with $\eta_{\alpha}$ of dimension $\mathrm{time}^{\alpha
-2}$. Its MSD grows like $\langle x^2(t)\rangle=2K_\alpha t^\alpha$
with $K_\alpha=[k_BT/(\Gamma(1+\alpha)m\eta_{\alpha})]$, which shows that
the FLE (\ref{fle}) describes a subdiffusion process. According to the
$\delta$-function property $\lim_{\alpha\to1^-}[|\tau|^{-\alpha}/\Gamma(1-
\alpha)] = 2\delta(\tau)$ \cite{gelfand1964}, the Markovian limit of this
description is obtained with $\langle\zeta_{1}(t)\zeta_{1}(t')\rangle=2[k_B
T\eta_1/m]\delta(t-t')$, which corresponds to the overdamped Langevin
equation (\ref{LE1}).

In this paper, we consider another variant of the Langevin equation that we
call fractional Langevin equation far from equilibrium (FLEFE),  
\begin{equation}\label{limCLE}
\frac{d^\alpha x(t)}{dt^\alpha}=\sqrt{2K_{\alpha}}\xi(t),
\end{equation}
in which the parameter is defined for all  $\alpha>0$, $\xi(t)$ is white
Gaussian noise as defined above and the fractional derivative is to be
interpreted in the Caputo sense (\ref{caputo-derivative}). Hereafter,
we show that the statistical quantities of the FLEFE exist for $\alpha>1/2$.
The three Langevin equations, i.e., the FBM (\ref{MNFBM}), the FLE (\ref{fle})
and the FLEFE (\ref{limCLE}), constitute a set of generic Langevin models
that account for memory effects, but in a different way: in the FBM case,
the particle is driven by an external random force exhibiting power-law
correlations;  in the FLE and FLEFE cases, the memory effects arise from
the viscoelastic properties of the media; the FLE obeys the FDT, but the
FBM and FLEFE do not.

The FLEFE was introduced by Eab and Lim~\cite{eab2011} who also derived the
exact solution which we reproduce in Sec. \ref{section4}. In particular,
with zero initial conditions the solution is reduced to the Riemann-Liouville
integral representation of FBM (RL-FBM) proposed by L{\'evy} \cite{levy1953},
which is also known as FBM II in the literature \cite{marinucci1999}. Moreover,
in Ref.~\cite{lim2001} Lim discussed ensemble averages such as the MSD and the
mean-squared increment (MSI) of the RL-FBM in the domain $1/2<\alpha<3/2$
(which as we will see below, corresponds to the Hurst exponent regime
$0<H<1$). Lim demonstrated that RL-FBM lacks the stationary property of
the increments, unlike the standard FBM, which has stationary increments
as described by Eq.~(\ref{MNFBM}). Instead, they identified an asymptotic
stationarity of the increments in the long-time behavior of RL-FBM indicating
that in this domain of the exponent $\alpha$, RL-FBM is asymptotically
ergodic at long times.

Here, we aim to develop the theory of FLEFE motion, Eq.~(\ref{limCLE}), for all
$\alpha>1/2$ beyond the standard Hurst exponent regime $0<H <1$, by focusing
on measurable quantities including the MSD, the time-averaged MSD (TAMSD),
the MSI and the autocovariance function (ACF) (i.e., the autocovariance of the
increments). We point the attention of the reader to the nonequivalence of the
generic definitions of the MSD and the mean TAMSD which leads to a surprising
spurious nonergodicity \cite{mardoukhi2020} in the regime $1/2<\alpha<3/2$,
and reveal a distinct coincidence between the mean TAMSD and the MSI in
the long-time limit. In view of the nonstationarity of the increments of
FLEFE motion, it is therefore a natural question to explore aging effects,
i.e., the explicit dependence of physical observables on the time span $t_a$
between the original system preparation and the start of the recording of
the particle motion.

The paper is organized as follows. In Sec.~\ref{section2}, we recall the
fundamental concepts of diffusion processes we plan to discuss in this
paper. In Sec.~\ref{section3}, we give a reminder of the characteristic
properties of two major stochastic processes, FBM and FLE motion. In
Sec.~\ref{section4}, we discuss the results for RL-FBM, as the solution
of FLEFE with zero initial condition. Specifically, we focus on ergodic
and asymptotically stationary properties by using the MSD, the MSI and
the mean TAMSD for all $\alpha>1/2$. In Sec.~\ref{section5} we derive the
form of the $(n+1)$th order MSI, and discuss its stationary property. In
Sec.~\ref{section6}, the aging effects for all $\alpha>1/2$ on the MSD and
mean TAMSD are considered. In Sec.~\ref{section7}, we summarize and discuss
our results.

\section{Statistical characteristics of diffusion processes}
\label{section2}

To quantify the averaged diffusive behavior of tracer particles, the
conventional measurable quantity is the MSD, defined by averaging over
the ensembles of trajectories $x_i(t)$ at time $t$ with respect to each
trajectory's initial position of the diffusing particles such as
\begin{equation}
\langle x^2(t)\rangle=\frac{1}{N}\sum_{i=1}^N (x_i(t)-x_i(0))^2 .
\end{equation}
Here $N$ is the total number of trajectories. The MSI, qualifying the increment
of displacement during the lag time $\Delta$ starting at physical time $t$,
is defined as the mean-squared of the increment in the form \cite{gikhman2004}
\begin{equation}
\label{ST}
\langle x^2_{\Delta}(t)\rangle=\langle[x(t+\Delta)-x(t)]^2\rangle.
\end{equation}
The MSI is equal to the MSD $\left<x^2(\Delta)\right>$ if the process has
stationary increments. The definition of the MSI is identical to the structure
function originally introduced by Kolmogorov and Yaglom in their works on
locally homogeneous and isotropic turbulence \cite{yaglom1953,yaglom1955,
yaglom1987, kolmogorov1941a, kolmogorov1941b}.

Alternatively, the diffusion of an individual particle can be quantified from a
single particle trajectory $x(t)$ via the TAMSD \cite{barkai2012, metzler2014}
\begin{equation}\label{TAMSD}
\overline{\delta^2(\Delta)}=\frac{1}{T-\Delta} \int_0^{T-\Delta}\left[x(t^\prime+\Delta)-x(t^\prime)\right]^2 d t^\prime,
\end{equation}
where $T$ is the length of the time series (measurement time) and $\Delta$
is the lag time. One can get the mean TAMSD  by averaging over an ensemble
of $N$ individual trajectories in the form
\begin{equation}\label{meanTAMSD}
\big<\overline{\delta^2(\Delta)}\big>=\frac{1}{N} \sum_{i=1}^N \overline{\delta_{ i}^2(\Delta)}.
\end{equation}
The TAMSD is typical used to evaluate few but long time series garnered in single particle tracking experiments, e.g., in biological cells, of geo-tagged larger animals, or of financial time series \cite{manzo2015, barkai2012, nathan2022, vilk2024}.

The concept of ergodicity which we will consider below relies on the
MSD-to-TAMSD equivalence in the limit of long trajectories and short lag
times, see \cite{he2008, metzler2014}, e.g.,
\begin{equation} \label{ergodic-relation}
\lim \limits_{\Delta/T \to 0} \overline{\delta^2(\Delta)} =\left\langle x^2(\Delta)\right\rangle.
\end{equation}
A stochastic process with stationary increment, for instance, BM and FBM,
is obviously ergodic according to this definition. Weakly nonergodic
models of anomalous diffusion such as continuous time random walks with
scale-free waiting times are easily distinguished from ergodic diffusion
models by applying tests \cite{meroz2013}, for instance the p-variation
test \cite{magdziarz2009} and the moving average vs ensemble average test
\cite{lubelski2008}. Numerous stochastic processes reveal weak ergodicity
breaking that violates the equivalence of the MSD and TAMSD. Starting
with the work of Bouchaud \cite{bouchaud1992}, there has been growing
interest in weak ergodicity breaking. A particular case is called ultraweak
ergodicity breaking \cite{bouchaud1992, godec2013}: here the MSD deviates
from the TAMSD only in the prefactor, albeit they have the same scaling
exponent. The work by Mardoukhi et al. \cite{mardoukhi2020} reported that
the Ornstein-Uhlenbeck process, known as a stationary and ergodic process,
leads to spurious non-ergodicity due to the failure of equivalence of
the generic MSD (depending on the initial condition $x(0)$) and TAMSD in
Eq.~(\ref{ergodic-relation}). The authors suggested that one should compare
the MSI and the TAMSD, a suggestion that we elaborate on in this paper.

In a nonstationary setting, the origin of time can no longer be chosen
arbitrarily. This raises the question of aging, that is, the explicit
dependence of physical observables on the time span $t_a$ between the
original preparation of the system and the start of the recording of
data. Traditionally, aging is considered as a key property of glassy systems
\cite{donth2001}. For an aged process, in which we measure the MSD starting
from the aging time $t_a$ until time $t$, the aged MSD is defined in the form
\begin{equation}\label{agedMSD}
\big< x^2(t)\big>_a=\big<\left[x(t_a+t)-x(t_a)\right]^2\big>,
\end{equation}
which is similar to the MSI in Eq.~(\ref{ST}). For a nonaged process with
$t_a=0$ the standard MSD is recovered, as it should. In the aged process,
the aged MSD (\ref{agedMSD}) is reduced by the amount accumulated until time
$t_a$, at which the measurement starts.

For an aged process originally initiated at $t=0$ and measured from $t_a$
for the duration (measurement time) $T$, the aged TAMSD is defined in the
form \cite{schulz2013, schulz2014}
\begin{equation}\label{agingTAMSD}
\overline{\delta_a^2(\Delta)}=\frac{1}{T-\Delta} \int_{t_a}^{T+t_a-\Delta}\left[x\left(t^{\prime}+\Delta\right)-x\left(t^{\prime}\right)\right]^2 d t^{\prime},  
\end{equation}
as a function of the lag time $\Delta$ and the aging time $t_a$. Averaging
over an ensemble of $N$ individual trajectories in the form
\begin{equation}
\big<\overline{\delta_a^2(\Delta)}\big>=\frac{1}{N} \sum_{i=1}^N \overline{\delta_{a, i}^2(\Delta)}
\end{equation}
defines the mean aged TAMSD.

\section{Statistical properties of FBM and FLE}
\label{section3}

FBM and FLE are important processes to describe anomalous diffusion, the
related quantities for these two major anomalous diffusion models are widely
studied \cite{schwarzl2017, deng2009, wang2020}. FBM and FLE have a different
physical nature, albeit both models share many common features. Here we
briefly recall the statistical properties of FBM and FLE.

\subsection{FBM}

The formal solution of FBM in Eq.~(\ref{MNFBM}) is
\begin{equation}
x(t)=\sqrt{2 K_\alpha} \int_0^t \xi_\alpha\left(t^{\prime}\right) d t^{\prime}, 
\end{equation}
in which the anomalous diffusion exponent range is $0<\alpha\le 2$.

Its ACF is \cite{mandelbrot1968}
\begin{equation}
\langle x(t_1)x(t_2)\rangle=K_\alpha\left(t_1^{\alpha}+t_2^{\alpha}-|t_2-t_1|
^{\alpha}\right) .
\end{equation}
Using the ACF, we find that the MSD, MSI, as well as the mean TAMSD obey
the same power-law \cite{deng2009,metzler2014}
\begin{equation}
\langle x^2_{\Delta}(t)\rangle=\langle x^2(\Delta)\rangle=\left<\overline{
\delta^2(\Delta)}\right>=2K_\alpha\Delta^\alpha,    
\end{equation}
which implies that FBM is an ergodic process.

\subsection{FLE}

The solution of the overdamped FLE in Eq.~({\ref{fle}}) is given by
\cite{molina2018}
\begin{equation}
x(t)=\frac{1}{\Gamma(\alpha)\eta_{\alpha}}\int_0^t(t-t')^{\alpha-1}\zeta
_\alpha\left(t'\right)dt',
\end{equation} 
in which the exponent $\alpha$ is defined in the range $(0, 1]$, corresponding
to normal diffusion and subdiffusion. Its ACF is \cite{molina2018}
\begin{equation}
\langle x(t_1)x(t_2)\rangle=K_\alpha\left(t_1^{\alpha}+t_2^{\alpha}-|t_2-t_1|^{
\alpha}\right) .
\end{equation}
with the generalized diffusion coefficient $K_\alpha=k_BT/[\Gamma(1+
\alpha)m\eta_{\alpha}]$. Thus we immediately arrive at the MSI, MSD, and mean
TAMSD with the scaling behaviors \cite{molina2018,deng2009}
\begin{equation}
\langle x^2_{\Delta}(t)\rangle=\langle x^2(\Delta)\rangle=\left<\overline{
\delta^2(\Delta)}\right>=2K_\alpha\Delta^{\alpha},
\end{equation} 
indicating that the FLE process FLE (\ref{fle}) is also a ergodic.

\section{FLEFE and RL-FBM}
\label{section4}

In what follows, we consider the FLEFE (\ref{limCLE}), in which the exponent
is defined for arbitrary $\alpha>0$, and the initial conditions are given
by $x^{(k)}(0)$, $k=0,\ldots,[\alpha]$, where $[\,\cdot\,]$ denotes the integer
part of a real-valued positive number. To emphasize the remarkable difference
between the FLEFE and the FLE, we first note that in the FLEFE, the random force
is represented by white Gaussian noise, breaking the FDT relation, and second,
that the FLEFE motion is well defined for arbitrary exponent $\alpha>0$.

The solution of Eq.~(\ref{limCLE}) was given as \cite{eab2011}
\begin{equation}
\label{GCLE}
x(t)=a(t)+\sqrt{2K_{\alpha}}\int_0^t\frac{(t-t')^{\alpha-1}}{\Gamma(\alpha)}\xi(t')
dt',
\end{equation}
in which the first part solely depends on the initial conditions, $a(t)=\sum_{k=0}
^{[\alpha]}t^{k}x^{(k)}(0)/\Gamma(k+1)$, and the second part is
identical to L{\'e}vy's formulation of FBM (RL-FBM) with Hurst exponent $H=\alpha
-1/2$ \cite{mandelbrot1968}. 

Without loss of generality, we here introduce the new process $x(t)\equiv x(t)
-\langle x(t)\rangle$, such that $\langle x(t)\rangle=0$. Equivalently, if one
assumes that all initial conditions are zero, i.e., $a(t)=0$, the solution of
the FLEFE (\ref{GCLE}) is equivalent to the RL-FBM \cite{lim2001}
\begin{equation}
\label{solution}
x(t)=\sqrt{2K_{\alpha}}\int_0^t\frac{(t-t')^{\alpha-1}}{\Gamma(\alpha)}\xi
(t')dt'.
\end{equation}
RL-FBM was first introduced by L{\'e}vy \cite{levy1953} and defined in terms of
a Riemann-Liouville fractional integral with initial condition at $t=0$. In
contrast to the equilibrated FBM developed by Mandelbrot and van Ness, L{\'e}vy's
RL-FBM is a stochastic process with non-stationary increments. This point gives
rise to interesting differences between FBM and FLEFE, as we now explore.

\subsection{MSD}

The ACF of FLEFE motion is given by (supposing that $t_1\geqslant t_2$)
\begin{equation}
\label{tpc}
\langle x(t_1)x(t_2)\rangle=\frac{2K_{\alpha}t_2^{\alpha}t_1^{\alpha-1}}{
\alpha\Gamma(\alpha)^2}\times{}_{2}F_{1}\left(1-\alpha, 1; \alpha+1; \frac{t_2}{t_1}\right),
\end{equation}
where 
\begin{equation}
{}_2F_1(a,b;c;z)=\frac{\Gamma(c)}{\Gamma(b)\Gamma(c-b)}\int_0^1\frac{t^{b-1}
(1-t)^{c-b-1}}{(1-tz)^a}dt
\end{equation}
is the hypergeometric function. By taking $t_1=t_2=t$ in the two-point correlation
function (\ref{tpc}) and using the formula \cite{abramowitz1964}
\begin{equation}
{}_2F_1(a,b;c;1)=\frac{\Gamma(c)\Gamma(c-a-b)}{\Gamma(c-a)\Gamma(c-b)}, \quad
\mathbb{R}(c)>\mathbb{R}(a+b),
\end{equation}
one can obtain the MSD \cite{eab2011}
\begin{equation}
\label{CaputoMSD}
\langle x^2(t)\rangle=\frac{2K_{\alpha}}{(2\alpha-1)\Gamma(\alpha)^2}t^{2\alpha-1}, 
\end{equation}
and thus one can see that the MSD is meaningful for all $\alpha>1/2$. It is worth
noting that this requirement for the exponent $\alpha$ in the FLEFE differs from,
e.g., those in SBM and FBM \cite{wei2023}.

\subsection{MSI}
\label{sectionstrfun}

Lim derived the MSI of RL-FBM in the form \cite{lim2001}
\begin{equation}
\label{increment11}
\langle x^2_{\Delta}(t)\rangle=\langle[x(t+\Delta)-x(t)]^2\rangle =\frac{2
K_{\alpha}}{\Gamma(\alpha)^2}\Delta^{2\alpha-1}\Bigg\{L_\alpha\left(\frac{t}{
\Delta}\right)+\frac{1}{2\alpha-1}\Bigg\},
\end{equation}
where the integral $L_\alpha\left(z\right)$ is given by 
\begin{equation}
L_\alpha(z)=\int_0^z\left[(1+s)^{\alpha-1}-s^{\alpha-1}\right]^2ds.
\end{equation}

In explicit form the MSI of RL-FBM can be expressed via the hypergeometric
function \cite{abramowitz1964} or the $H$-function \cite{prudnikov1990} (see
the derivation of Eq.~(\ref{a4}) in App.~\ref{AppendixA}) 
\begin{equation}
\label{increment01}
\langle x^2_{\Delta}(t)\rangle=\frac{2K_{\alpha}}{(2\alpha-1)\Gamma(\alpha)^2}
[(t+\Delta)^{2\alpha-1}+t^{2\alpha-1}]-\frac{4K_{\alpha}t^{\alpha}(t+\Delta)^{
\alpha-1}}{\alpha\Gamma(\alpha)^2}{}_2F_1\left(1-\alpha,1;1+\alpha;\frac{t}{t+
\Delta}\right).
\end{equation}
We note here that both expressions (\ref{increment11}) and (\ref{increment01})
of the MSI are valid for all $\alpha>1/2$ and underline the non-stationary
nature of RL-FBM. In particular, when $\alpha$ approaches $1$, the hypergeometric
function ${}_2F_1(1-\alpha,1;1+\alpha;t/{(t+\Delta)})=1$ and thus $\langle x^2_
{\Delta}(t)\rangle=2K_1\Delta$. 

For sufficiently short times $t\ll\Delta$, the integral $L_\alpha(t/\Delta)\approx
0$ in Eq.~(\ref{increment11}), and the MSI reads
\begin{equation}
\label{STshort}
\langle x^2_{\Delta}(t)\rangle\sim\frac{2K_\alpha}{(2\alpha-1)\Gamma(\alpha)
^2}\Delta^{2\alpha-1},
\end{equation}
which is identical to the MSD (\ref{CaputoMSD}) of RL-FBM. 

We proceed to discuss the MSI in the long time limit $t\gg\Delta$ in the three
relevant different regimes for the order $\alpha$ of the fractional derivative
in the FLEFE (RL-FBM). We arrive at the same approximation for the MSI by using
two approaches: The first one, which is presented in the main text, uses the
approximation of the integral Eq.~(\ref{increment11}), while the second one
applies a series expansion of the hypergeometric function or the Fox
$H$-function in Eq.~(\ref{increment01}); the latter is shown in
App.~\ref{AppendixA}.

\subsubsection{Case $1/2<\alpha<3/2$}

In the long time limit $t\gg\Delta$, the integral in Eq.~(\ref{increment11})
converges as \cite{balcerek2022}
\begin{equation}
L_\alpha\left(\frac{t}{\Delta}\right)+\frac{1}{2\alpha-1}\approx\int_0^{\infty}
\left[(1+s)^{\alpha-1}-s^{\alpha-1}\right]^2ds+\frac{1}{2\alpha-1}=\frac{
\Gamma(\alpha)^2}{\Gamma(2\alpha)|\cos(\pi\alpha)|},
\end{equation} 
and thus we arrive at the the stationary MSI approximated as \cite{lim2001}
\begin{equation}
\label{increment-long-time1}
\langle x^2_{\Delta}(t)\rangle\sim\frac{2K_{\alpha}}{\Gamma(2\alpha)|\cos(\pi
\alpha)|}\Delta^{2\alpha-1},
\end{equation}
which depends solely on the lag time $\Delta$. The MSI becomes stationary in
the long-time regime $t\gg\Delta$ and has the same anomalous diffusion exponent
as the MSD (\ref{CaputoMSD}) of RL-FBM, but deviates from the MSD by a constant
prefactor.

\subsubsection{Case $\alpha>3/2$}

For $\alpha> 3/2$, the integral in Eq.~(\ref{increment11}) asymptotically reads
\begin{equation}
L_\alpha\left(\frac{t}{\Delta}\right)=\int_0^{t/\Delta}s^{2\alpha-2}\left[(1+s^{
-1})^{\alpha-1}-1\right]^2ds\sim\frac{(\alpha-1)^2}{2\alpha-3}\left(\frac{t}{
\Delta}\right)^{2\alpha-3},
\end{equation} 
and thus the MSI is given by 
\begin{equation}
\label{mTAMSD}
\langle x^2_{\Delta}(t)\rangle\sim\frac{2(\alpha-1)^{2}K_{\alpha}}{(2\alpha-3)
\Gamma(\alpha)^2}t^{2\alpha-3}\Delta^2.
\end{equation}
In this range of $\alpha$ in the FLEFE, the MSI depends ballistically on the lag
time $\Delta$ and keeps a dependence on the measured time, scaling as $t^{2\alpha
-3}$.

\subsubsection{Case $\alpha=3/2$}

Apparently, it follows from Eqs.~(\ref{increment-long-time1}) and (\ref{mTAMSD})
that the case $\alpha=3/2$ needs to be discussed separately. In this case the
MSI (\ref{ST}) can be rewritten as
\begin{equation}
\label{ST32}
\langle x^2_{\Delta}(t)\rangle=\frac{8K_{3/2}}{\pi}\left[L_{3/2}\left(\frac{t}{
\Delta}\right)+\frac{1}{2}\right]\Delta^2,
\end{equation} 
where the integral $L_{3/2}(z)$ can be evaluated explicitly as
\begin{equation}
\label{integral1}
L_{3/2}(z)=\int_0^z\left(\sqrt{1+s}-\sqrt{s}\right)^2ds=z^2+z+\frac{1}{2}\left[
\sinh^{-1}(\sqrt{z})-\sqrt{z}\sqrt{z+1}(2z+1)\right].
\end{equation}
When $t\gg\Delta$, the integral can be approximated by the leading term 
\begin{equation}
\label{eq42}
L_{3/2}\left(\frac{t}{\Delta}\right)\sim\frac{1}{4}\ln\left(\frac{t}{\Delta}
\right).
\end{equation}
Therefore, inserting Eq.~(\ref{eq42}) into the Eq.~(\ref{ST32}) yields the MSI
with the dominant term
\begin{equation}
\langle x^2_{\Delta}(t)\rangle\sim\frac{2K_{3/2}}{\pi}\Delta^2\ln\left(
\frac{t}{\Delta}\right),
\end{equation}   
which also implies nonstationarity in the sense that the $t$-dependence remains.  

The MSI of RL-FBM in the long time limit $t\gg\Delta$ for all $\alpha>1/2$ can
thus be summarized as
\begin{equation}
\label{STlong}
\langle x_{\Delta}^2(t)\rangle\sim\left\{\begin{array}{ll}
\displaystyle\frac{2K_\alpha}{\Gamma{(2\alpha)}|\cos(\pi\alpha)|}\Delta^{2\alpha-1},
&\quad 1/2<\alpha<3/2\\[0.32cm]
\displaystyle\frac{2K_{3/2}}{\pi}\Delta^2\ln\left(\frac{t}{\Delta}\right),
&\quad\alpha=3/2\\[0.32cm]
\displaystyle\frac{2(\alpha-1)^2K_{\alpha}}{(2\alpha-3)\Gamma(\alpha)^2}t^{2\alpha
-3}\Delta^2, & \quad \alpha>3/2\end{array}\right..
\end{equation} 
As mentioned above, another approach  to calculate the MSI is presented in
App.~\ref{AppendixA}.

Unlike the FBM (\ref{MNFBM}), which has stationary increments, the RL-FBM has
nonstationary increments depending on the measured time $t$, see the explicit
form of the MSI (\ref{increment01}) of RL-FBM. In addition, we observe that the
MSI tends to become approximately stationary at long times when $1/2<\alpha<3/2$,
whereas for $\alpha\ge3/2$ the MSI remains nonstationary.

\subsection{TAMSD}

According to the definition (\ref{TAMSD}) of the TAMSD, the mean TAMSD of RL-FBM
can be derived as (see App.~\ref{AppendixB})
\begin{eqnarray}
\label{analyticalTAMSD1}
\nonumber
\left<\overline{\delta^2(\Delta)}\right>&=&\frac{1}{T-\Delta}\int_0^{T-\Delta}
\langle[x(t+\Delta)-x(t)]^2\rangle dt\\
&=&\frac{K_\alpha}{\alpha(2\alpha-1)\Gamma(\alpha)^2}\frac{T^{2\alpha}-\Delta
^{2\alpha}}{T-\Delta}+\frac{K_\alpha}{\alpha(2\alpha-1)\Gamma(\alpha)^2}(T-
\Delta)^{2\alpha-1}-I_{\alpha}(\Delta,T),
\end{eqnarray}
where
\begin{equation}
\label{Ia}
I_{\alpha}(\Delta,T)=\frac{4K_\alpha}{\alpha\Gamma(\alpha)^2}\frac{1}{T-\Delta}
\int_0^{T-\Delta}(t+\Delta)^{\alpha-1}t^\alpha\times{}_2F_1\left(1-\alpha,1;1+
\alpha;\frac{t}{t+\Delta}\right)dt. 
\end{equation}
In what follows we discuss the approximations of the mean TAMSD at long times, $T\gg\Delta$, in the different regimes of $\alpha$.

\subsubsection{Case $1/2<\alpha< 3/2$}

From the derivation of Eq.~(\ref{b10}) in App.~\ref{AppendixB} we obtain the mean
TAMSD of RL-FBM in Eq.~(\ref{analyticalTAMSD1}) in the long time limit $T\gg\Delta$
in the form
\begin{equation}
\label{TAMSD1}
\left<\overline{\delta^2(\Delta)}\right>\sim\frac{2K_{\alpha}}{\Gamma(2\alpha)
|\cos(\pi\alpha)|}\Delta^{2\alpha-1}.
\end{equation}
The mean TAMSD in this range of $\alpha$ deviates from the MSD (\ref{CaputoMSD})
in the prefactor but has the same form as the MSI (\ref{increment-long-time1}).

\subsubsection{Case $\alpha>3/2$}

According to the derivation of Eq.~(\ref{b11}) in App.~\ref{AppendixB} we obtain
the mean TAMSD (\ref{analyticalTAMSD1}) of RL-FBM in the long time limit $T\gg
\Delta$ as
\begin{equation}
\label{T42}
\left<\overline{\delta^2(\Delta)}\right>\sim\frac{(\alpha-1)K_{\alpha}}{(2\alpha
-3)\Gamma(\alpha)^2}T^{2\alpha-3}\Delta^2.
\end{equation}
In this case the mean TAMSD depends ballistically on the lag time $\Delta$ and
keeps a dependence on the measurement time $T$ with the scaling $T^{2\alpha-3}$.

\subsubsection{Case $\alpha=3/2$}

Using the explicit power-logarithmic expansion of the $H$-function for $\alpha
=3/2$ as shown in Eq.~(\ref{b12}), one can obtain the approximation of the TAMSD
(see the derivation of Eq.~(\ref{b15}) in App.~\ref{AppendixB} for details) when
$T/\Delta\gg 1$,
\begin{equation}
\left<\overline{\delta^2(\Delta)}\right>\sim\frac{2K_{3/2}}{\pi}\Delta^2\ln
\left(\frac{T}{\Delta}\right).
\end{equation}

In summary, the mean TAMSD of the RL-FBM in the three relevant regimes of
$\alpha$ reads
\begin{equation}
\label{TAMSDlong}
\left<\overline{\delta^2(\Delta)}\right>\sim\left\{\begin{array}{ll}
\displaystyle\frac{2K_{\alpha}}{\Gamma(2\alpha)|\cos(\pi\alpha)|}\Delta^{2
\alpha-1},\quad&1/2<\alpha< 3/2\\[0.32cm]
\displaystyle\frac{2K_{3/2}}{\pi}\Delta^2\ln\left(\frac{T}{\Delta}\right),&
\alpha=3/2\\[0.32cm]
\displaystyle\frac{(\alpha-1)K_{\alpha}}{(2\alpha-3)\Gamma(\alpha)^2}\Delta^2
T^{2\alpha-3},&\alpha>3/2\end{array}\right..
\end{equation} 

\begin{figure}
(a)\includegraphics[width=0.43\textwidth]{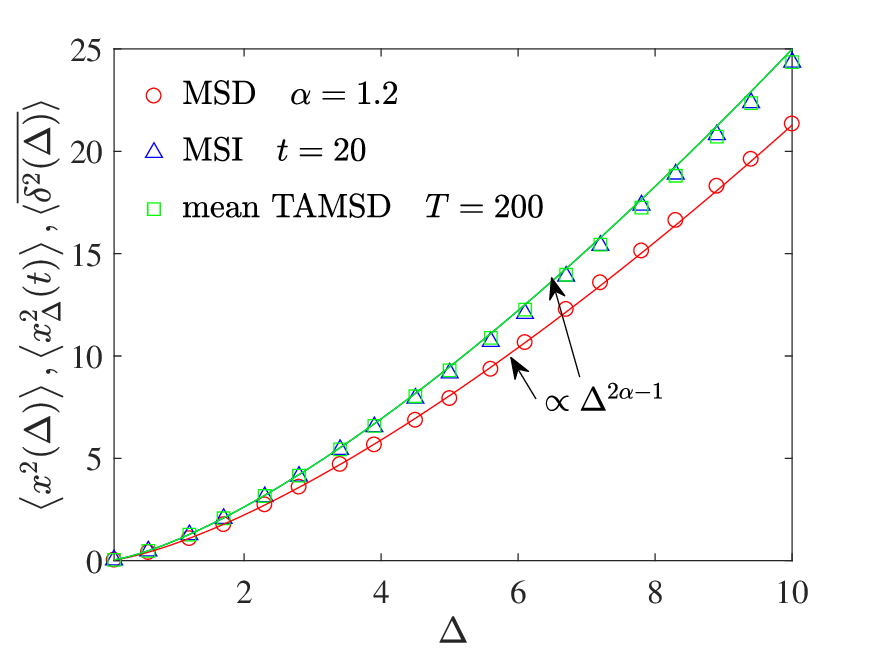}
(b)\includegraphics[width=0.43\textwidth]{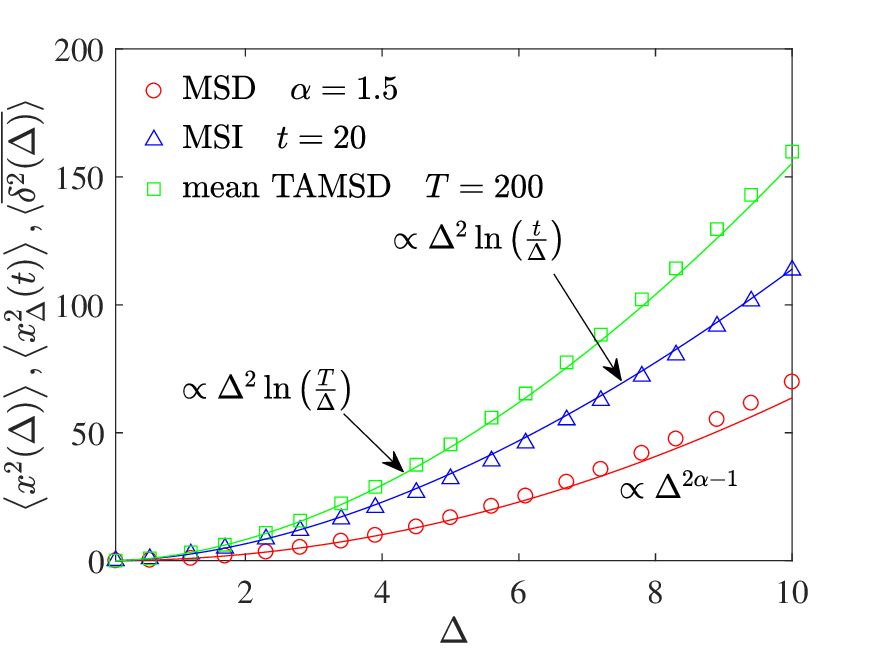}
(c)\includegraphics[width=0.43\textwidth]{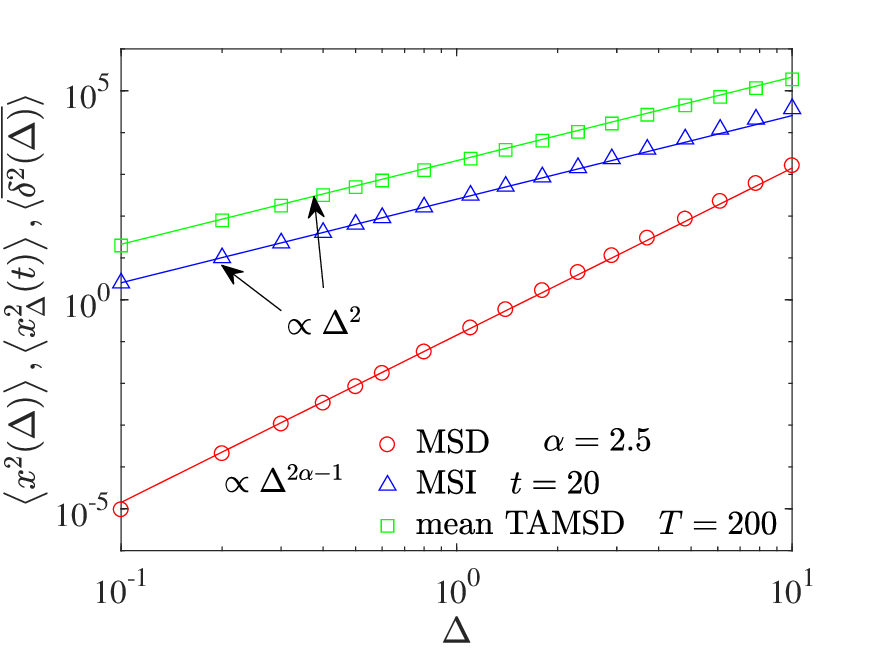}
\caption{Simulations (symbols) and analytical solutions (solid curves) for the
MSD (\ref{CaputoMSD}), MSI (\ref{increment11}), and mean TAMSD
(\ref{analyticalTAMSD1}) of RL-FBM for (a) $1/2<\alpha<3/2$, (b) $\alpha=3/2$, and
(c) $\alpha>3/2$. When $1/2<\alpha<3/2$, the mean TAMSD is identical to the MSI and
not to the MSD, leading to a spurious nonergodicity. When $\alpha\ge3/2$, RL-FBM
is nonergodic, and the mean TAMSD exhibits ballistic motion. Other parameters:
time step $dt=0.1$ and $K_{\alpha}=0.5$. The algorithm for the simulations is
presented in App.~\ref{AppendixC}.}
\label{Figure1}
\end{figure}

The results of our simulations for the moments (MSD, MSI, TAMSD) are shown in
Fig.~\ref{Figure1} for RL-FBM with different $\alpha$ along with the analytical
solutions (\ref{CaputoMSD}) for the MSD, the MSI (\ref{increment11}), and the mean
TAMSD (\ref{analyticalTAMSD1}). The mean TAMSD is identical to the MSI when $1/2<\alpha
<3/2$ for sufficiently long trajectories, while the discrepancies between the
mean TAMSD and the MSI or MSD is retained when $\alpha\ge3/2$. The
algorithm of the numerical approach to generate the trajectories of RL-FBM
(\ref{solution}) is presented in App.~\ref{AppendixC}.

\begin{figure}
(a)\includegraphics[width=0.43\textwidth]{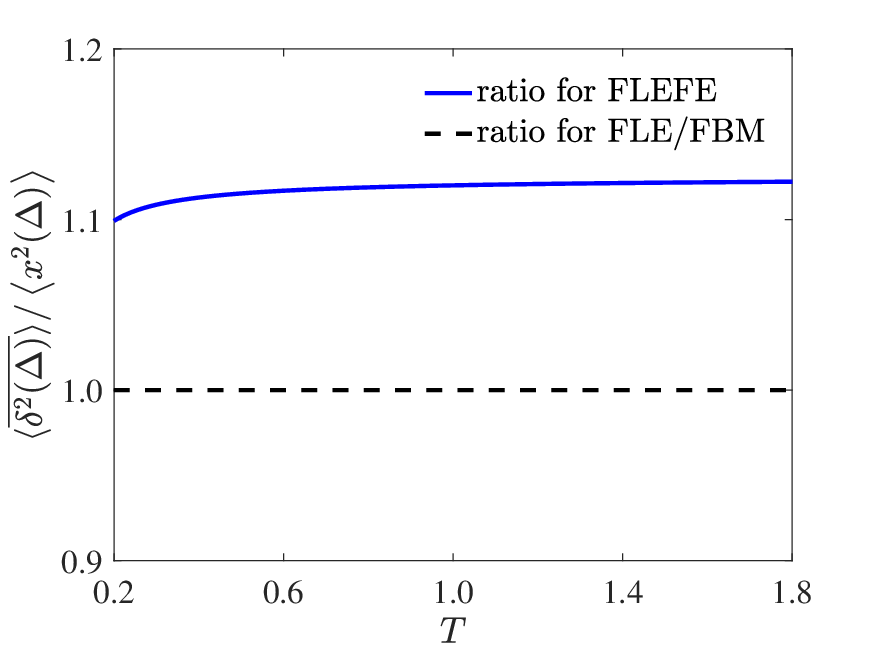}
(b)\includegraphics[width=0.43\textwidth]{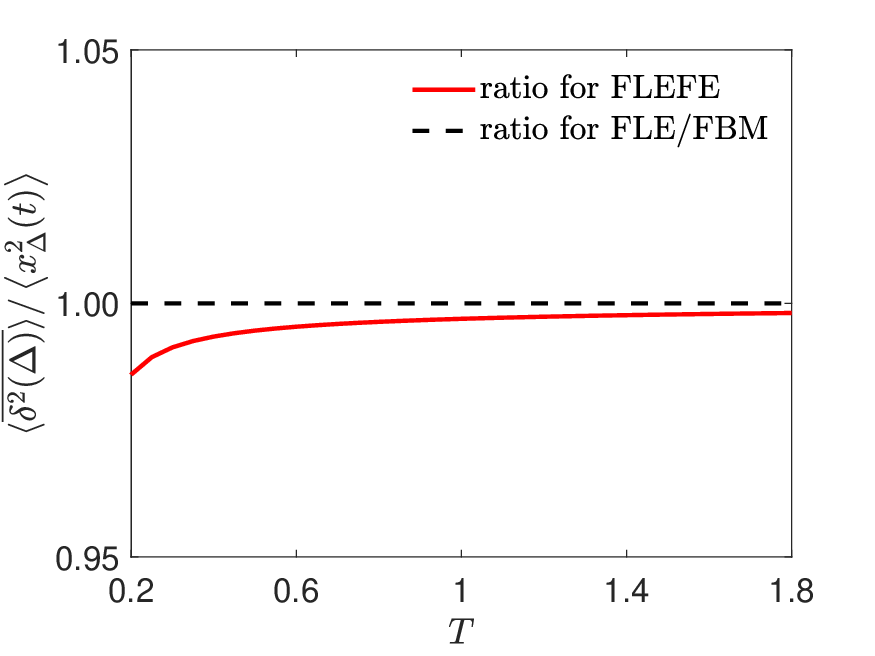}
(c)\includegraphics[width=0.43\textwidth]{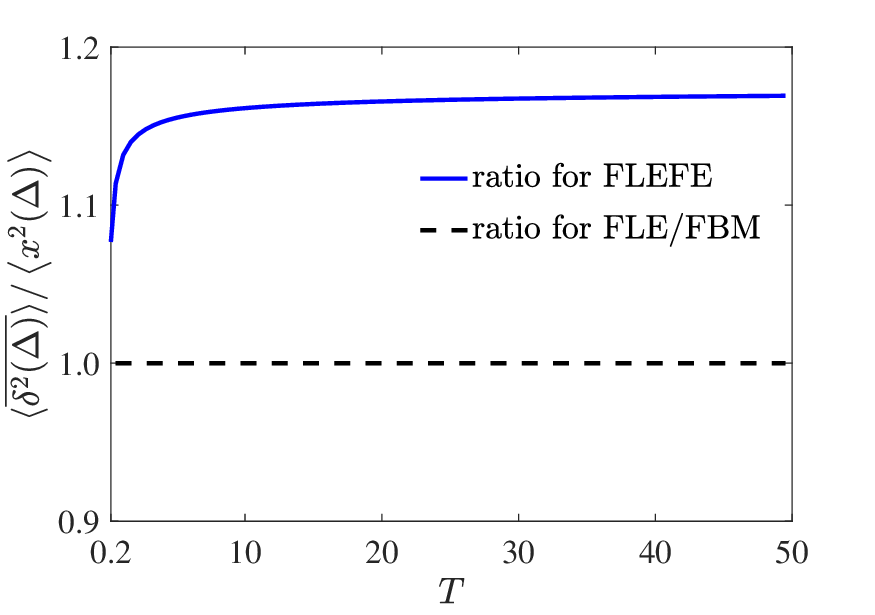}
(d)\includegraphics[width=0.43\textwidth]{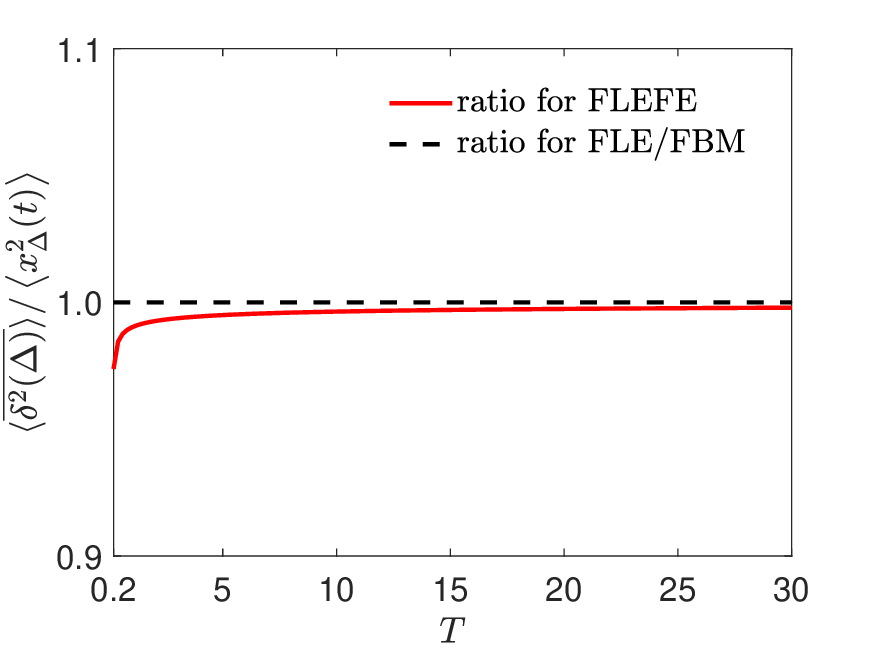}
\caption{Ratio $\left<\overline{\delta^2(\Delta)}\right>/\langle x^2_{\Delta}
(t)\rangle$ and $\left<\overline{\delta^2(\Delta)}\right>/\langle x^2(\Delta)
\rangle$ as function of the measurement time $T$ for different regimes. In (a)
and (b), $\alpha_1=0.8$, and in (c) and (d), $\alpha_2=1.2$. When $1/2<\alpha
< 3/2$, in the long time limit $T/\Delta\gg1$ the mean TAMSD of RL-FBM (FLEFE)
does not converge to its MSD but to the MSI in the strong aging limit. Here we
set $t=T/2$ for the MSI. In all panels we chose the lag time as $\Delta=0.1$;
the time step is $dt=0.1$, and $K_{\alpha}=0.5$.}
\label{Figure2}
\end{figure}

Figure~\ref{Figure2} shows the ratios of the mean TAMSD (\ref{analyticalTAMSD1})
to the MSD (\ref{CaputoMSD}) and the MSI (\ref{increment11}) with $1/2<\alpha
<3/2$, illustrating that at long times $T/\Delta\gg1$ the TAMSD of RL-FBM
converges to the MSI rather than the MSD.

In the context of single particle tracking experiments the conclusion about
ergodicity of the process (in the mean squared sense) is deduced from the
convergence of the mean TAMSD to the MSD for sufficiently long trajectories,
see Eq.~(\ref{ergodic-relation}) in \cite{metzler2014}. However, by its
construction, we show here that it is advisable to compare the mean TAMSD
(\ref{meanTAMSD}) with the MSI (\ref{ST}), rather. The MSD and MSI coincide
for processes with stationary increments, such as FBM and FLE motion. However,
this is not the case for processes with non-stationary increments. Taking RL-FBM
as an exemplary process one can see that the mean TAMSD (\ref{TAMSDlong}) and MSD
(\ref{CaputoMSD}) differ in the entire domain $\alpha>1/2$ even in the long
trajectory limit. This may lead to the conclusion about non-ergodicity of RL-FBM
in the entire domain of the fractional exponent. However, it is a spurious
non-ergodicity in the domain $1/2<\alpha<3/2$. Indeed, the increments of RL-FBM
in this regime become asymptotically stationary as follows from
Eq.~(\ref{increment-long-time1}), and the mean TAMSD converges to the MSI in the
limit of long trajectories, compare Eqs.~(\ref{increment-long-time1}) and
(\ref{TAMSD1}).

\section{Stationarity of the $(n+1)$th order MSI for RL-FBM}
\label{section5}
 
In the previous section, the MSI (\ref{STlong}) of RL-FBM was found to be
stationary at long times $t\gg\Delta$ in the regime $1/2<\alpha<3/2$, whereas
it is nonstationary when $\alpha\ge3/2$. A natural question is whether
stationarity is recovered in higher orders of the MSI of RL-FBM. The aim of
this section is to investigate the regime of the exponent $\alpha$ in which
the $n$th order MSI will restore the stationarity property. The $(n+1)$th
order increment is defined as the difference of the $n$th order increment
over a lag time $\tau_{n+1}$ \cite{schulz20131},
\begin{equation}
\Delta^{(n+1)}x(t;\tau_1,\ldots,\tau_{n+1})=\Delta^{(n)}x(t+\tau_{n+1};\tau_1,
\ldots,\tau_n)-\Delta^{(n)}x(t;\tau_1,\ldots,\tau_n),
\end{equation}
similar in spirit to the theory of $n$th order increments by Yaglom
\cite{yaglom1955,yaglom1953}.

\begin{figure}
\includegraphics[width=0.43\textwidth]{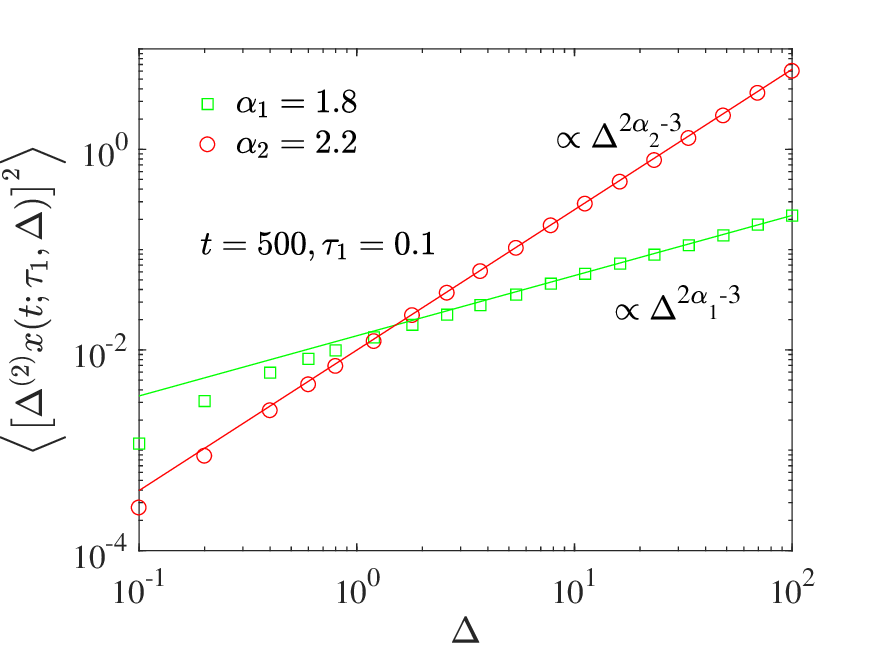}
\caption{Simulations (symbols) for the second order MSI of RL-FBM at long times
$t\gg\Delta+\tau_1$ for different $\alpha$. The theoretical result (\ref{delta11})
is represented by solid lines. Other parameters: $dt=0.1$, $\tau_1=0.1$, and  $K_{\alpha}=0.5$.}
\label{Figure3}
\end{figure}

According to this definition, the first and second order increments are given by
\begin{equation}
\Delta^{(1)}x(t;\tau_1)=x(t+\tau_1)-x(t)
\end{equation}
and
\begin{equation}
\Delta^{(2)}x(t;\tau_1,\tau_2)=\Delta^{(1)}x(t+\tau_2;\tau_1)-\Delta^{(1)}x(t;
\tau_1),
\end{equation}
where $\Delta^{(1)}x(t;\tau_1)$ is the first order increment of the displacement
and $\Delta^{(2)}x(t;\tau_1,\tau_2)$ is an increment of the first order
increment. With the approximation
\begin{equation}
\frac{\Delta^{(1)}x(t;\tau_1)}{\tau_1}=\frac{x(t+\tau_1)-x(t)}{\tau_1}\approx
\frac{dx(t)}{dt}
\end{equation}
at long times $t\gg\tau_1$ for the first order increment, one yields the integral
representation  
\begin{equation}
\label{delta1}
\Delta^{(1)}x(t;\tau_1)=\tau_1\frac{dx(t)}{dt}=\frac{\sqrt{2K_\alpha}}{\Gamma(
\alpha-1)}\tau_1\int_0^t(t-t')^{\alpha-2}\xi(t')dt'.
\end{equation}
We note that the integral representation of the first order increment is analogous
to that of the FLEFE (\ref{GCLE}) whose MSI was found to become stationary in the
regime $1/2<\alpha<3/2$ given by Eq.~(\ref{increment-long-time1}). Then, by a
similar computation, we obtain that the second order MSI in the regime $3/2<
\alpha<5/2$ takes on the form 
\begin{eqnarray}
\label{delta11}
\left<\left[\Delta^{(2)}x(t;\tau_1, \Delta)\right]^2\right>=
\left<\left[\Delta^{(1)}x(t+\Delta;\tau_1)-\Delta^{(1)}x(t;\tau_1)\right]^2\right>
\sim\frac{2K_{\alpha}\tau_1^2}{\Gamma(2\alpha-2)\cos(\pi\alpha)}\Delta^{2\alpha-3}
\end{eqnarray}
in the long time limit $t\gg\tau_{1}+\Delta$. This latter result is independent of
the measurement time $t$, and thus revels the stationarity of the second order
increment. The simulations of the second order MSI is shown in Fig.~\ref{Figure3},
showing perfect agreement with the analytical results.

More generally, the $(n+1)$th MSI 
\begin{eqnarray}
\label{delta1n}
\nonumber
\left<\left[\Delta^{(n+1)}x(t;\tau_1,\ldots,\tau_n,\Delta)\right]^2\right>
&=&\left<\left[\Delta^{(n)}x(t+\Delta;\tau_1,\ldots,\tau_n)-\Delta^{(n)}
x(t;\tau_1,\ldots,\tau_n)\right]^2\right>\\
&\sim&\frac{2K_{\alpha}(\tau_1\times\ldots\times\tau_n)^2}{\Gamma(2\alpha-2n)
|\cos(\pi(\alpha-n))|}\Delta^{2\alpha-2n-1}
\end{eqnarray}
restores stationarity in the regime $(2n+1)/2<\alpha<(2n+3)/2$.

\section{Aging effects}
\label{section6}

A distinct effect that can be probed for experimental data is aging, leading
to a possible dependence on the aging time $t_a$. We here analyze aging for
the MSD and TAMSD.

\subsection{Aging of the MSD}

As the aged MSD with aging time $t_a$ defined in Eq.~(\ref{agedMSD}) is analogous
to the MSI (\ref{increment11}) of RL-FBM with process time $t$, one can immediately
obtain the aged MSD. According to the results for the MSI of RL-FBM in
Eqs.~(\ref{STshort}) and (\ref{STlong}), we find the aged MSD for weak aging ($t_a
\ll t$),
\begin{equation}
\label{aged-msd-weak}
\langle x^2(t)\rangle_a\sim\frac{2K_\alpha}{(2\alpha-1)\Gamma(\alpha)^2}t^{2
\alpha-1}
\end{equation}
for all $\alpha>1/2$, which is identical to the nonaged MSD (\ref{CaputoMSD})
of RL-FBM.

For strong aging, $t_a\gg t$, the aged MSD for $1/2<\alpha<3/2$ is independent
of the aging time $t_a$, exhibiting the same power-law scaling as the weakly
aged MSD but with a different prefactor, whereas for $\alpha\ge3/2$, the aged
MSD has a ballistic scaling with the aging time $t_a$. We summarize the aged MSD
as
\begin{equation}
\label{aged-msd-strong}
\langle x^2(t)\rangle_a\sim\left\{\begin{array}{ll}
\displaystyle\frac{2K_\alpha}{\Gamma{(2\alpha)}|\cos(\pi\alpha)|}t^{2\alpha-1},
\quad & 1/2<\alpha<3/2\\[0.32cm]
\displaystyle\frac{2K_{3/2}}{\pi}t^2\ln\left(\frac{t_a}{t}\right),&\alpha=3/
2\\[0.32cm]
\displaystyle\frac{2(\alpha-1)^2K_{\alpha}}{(2\alpha-3)\Gamma(\alpha)^2}t^{2
\alpha-3}_at^2,&\alpha>3/2\end{array}\right..
\end{equation}

\subsection{Aging of the TAMSD}

According to the definition (\ref{agingTAMSD}) of the aged TAMSD, we obtain the
explicit aged mean TAMSD as
\begin{eqnarray}
\label{maTAMSD1}
\nonumber
\left<\overline{\delta_a^2(\Delta)}\right>&=&\frac{1}{T-\Delta}\int_{t_a}^{
T+t_a-\Delta}\left<\left[x(t'+\Delta)-x(t')\right]^2\right>dt'\\
&&\hspace*{-1.2cm}
=\frac{K_\alpha}{\alpha(2\alpha-1)\Gamma(\alpha)^2}\frac{(T+t_a)^{2\alpha}-(
t_a+\Delta)^{2\alpha}}{T-\Delta}+\frac{K_\alpha}{\alpha(2\alpha-1)\Gamma(\alpha)
^2}\frac{(T+t_a-\Delta)^{2\alpha}-{t_a}^{2\alpha}}{T-\Delta}+Q_{\alpha}(\Delta,
t_a,T).
\end{eqnarray}
Here $Q_{\alpha}(\Delta,t_a,T)$ is given via the hypergeometric function in the
form 
\begin{equation}
Q_{\alpha}(\Delta,t_a,T)=-\frac{4K_\alpha}{\alpha\Gamma(\alpha)^2}\frac{1}{T-
\Delta}\int_{t_a}^{T+t_a-\Delta}(t+\Delta)^{\alpha-1}t^\alpha{}_2F_1\left(1-
\alpha,1;1+\alpha;\frac{t}{t+\Delta}\right)dt.
\end{equation} 
One may yield approximations of the mean aged TAMSD from expansion of the
hypergeometric function or the $H$-function, analogous to the nonaged mean
TAMSD in App.~\ref{AppendixB}. Here we simply present the results for the
mean aged TAMSD of RL-FBM with different $\alpha$. 

Thus, for weak aging $t_a\ll T$ and $T\gg\Delta$ the results for the
asymptotic behaviors are summarized as 
\begin{equation}
\label{aged-tamsd-weak}
\left<\overline{\delta_a^2(\Delta)}\right>\sim\left\{\begin{array}{ll}
\displaystyle\frac{2K_{\alpha}}{\Gamma(2\alpha)|\cos(\pi\alpha)|}\Delta^{2
\alpha-1},\quad &1/2<\alpha<3/2\\[0.32cm]
\displaystyle\frac{2K_{3/2}}{\pi}\Delta^2\ln{\left(\frac{T}{\Delta}\right)},&
\alpha=3/2\\[0.32cm]
\displaystyle\frac{(\alpha-1)K_{\alpha}}{(2\alpha-3)\Gamma(\alpha)^2}\Delta^2
T^{2\alpha-3},&\alpha>3/2\end{array}\right.,
\end{equation} 
which coincides with the nonaged mean TAMSD (\ref{TAMSDlong}).

For strong aging $t_a\gg T$, the mean aged TAMSD has the same asymptotics as the
aged MSD (\ref{aged-msd-strong}) in the strong aging regime $t_a\gg t$, namely,
\begin{equation}
\label{aged-tamsd-strong}
\left<\overline{\delta_a^2(\Delta)}\right>\sim\left\{\begin{array}{ll}
\displaystyle\frac{2K_{\alpha}}{\Gamma(2\alpha)|\cos(\pi\alpha)|}\Delta^{2
\alpha-1},\quad&1/2<\alpha<3/2\\[0.32cm]
\displaystyle\frac{2K_{3/2}}{\pi}\Delta^2\ln{\left(\frac{t_a}{\Delta}\right)},&
\alpha=3/2\\[0.32cm]
\displaystyle\frac{2(\alpha-1)^2K_{\alpha}}{(2\alpha-3)\Gamma(\alpha)^2}\Delta^2
t_a^{2\alpha-3},&\alpha>3/2\end{array}\right.,
\end{equation}
from which ergodicity is restored in the regime $1/2<\alpha<3/2$, since in the
limit $t_a\gg T$ the collection of the increments $x(t+\Delta)-x(t)$ for $t\in
[t_a,t_a+T-\Delta]$ in the TAMSD (\ref{agingTAMSD}) changes only marginally and
is almost identical to $x(t_a+\Delta)-x(t_a)$.

Simulations of the aged MSD and the mean aged TAMSD for RL-FBM with different
$\alpha$ are shown in Fig.~\ref{Figure4}, demonstrating perfect agreement with
our theoretical results in the weak and strong aging regimes. For completeness
we also analyzed the aged ACF of RL-FBM increments, and the results are
presented in App.~\ref{AppendixD}.

\begin{figure}
(a)\includegraphics[width=0.43\textwidth]{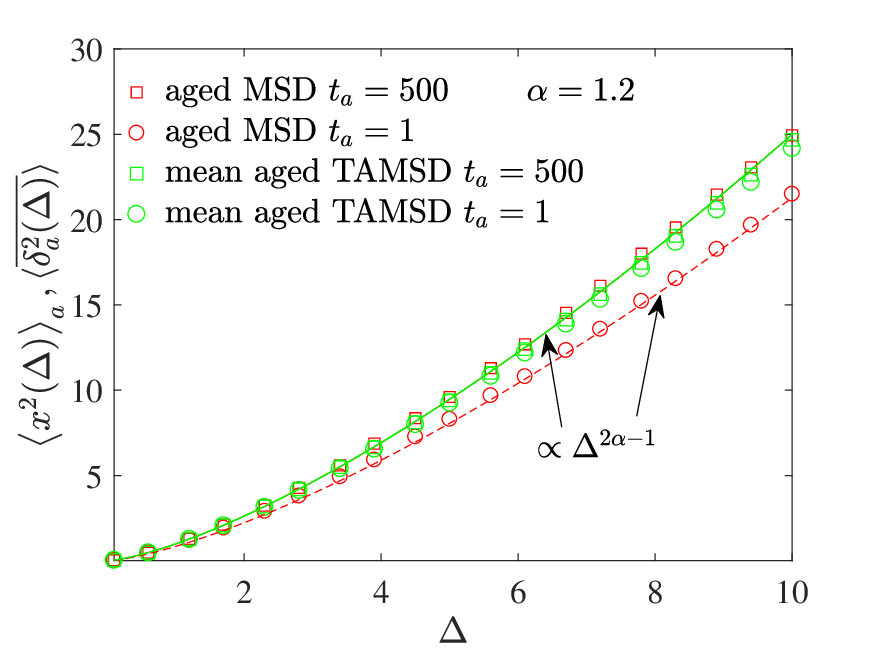}
(b)\includegraphics[width=0.43\textwidth]{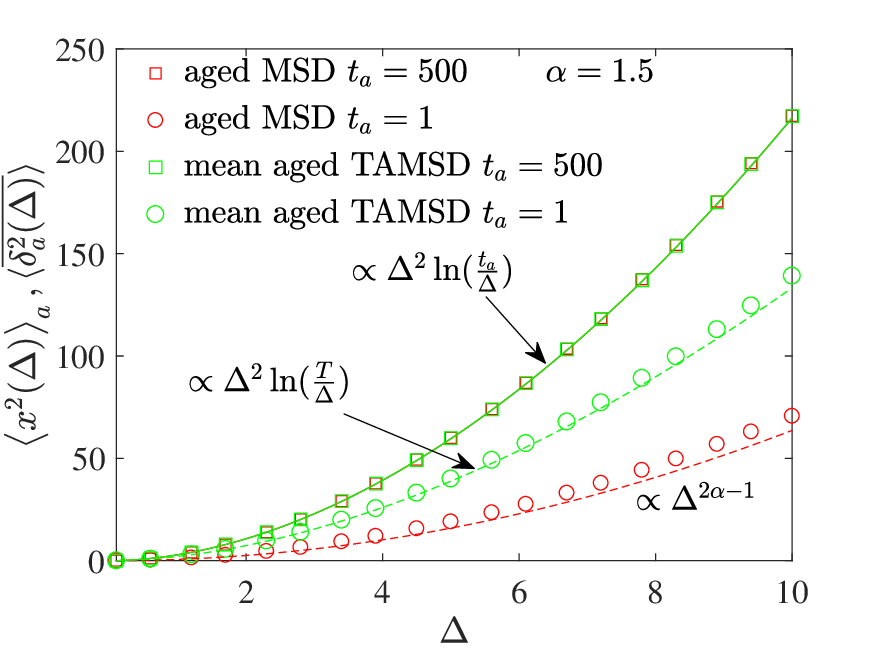}
(c)\includegraphics[width=0.43\textwidth]{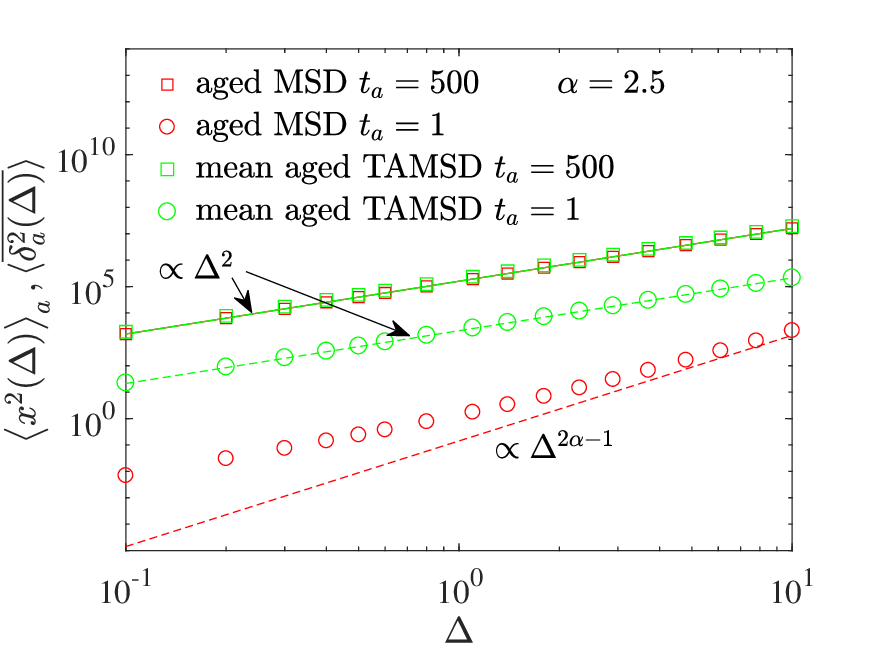}
\caption{Simulations and analytical results for the aged MSD and mean aged TAMSD
of RL-FBM for (a) $1/2<\alpha<3/2$, (b) $\alpha=3/2$ and (c) $\alpha>3/2$. The
theoretical weakly aged MSD (\ref{aged-msd-weak}), and the mean aged TAMSD
(\ref{aged-tamsd-weak}) are represented by dashed curves, while the strongly
aged MSD (\ref{aged-msd-strong}) and mean aged TAMSD (\ref{aged-tamsd-strong})
are represented by solid curves. Note that the (red and green) solid curves
overlap since ergodicity is restored in the strong aging case. Other parameters:
the simulation time step is $dt=0.1$, the measurement time is $T=100$, and
$K_{\alpha}=0.5$.}
\label{Figure4}
\end{figure}

\section{Conclusions}
\label{section7}

We introduced and discussed FLEFE motion, that provides a Langevin equation
based formulation as an alternative to the FBM and FLE approaches for systems
characterized by long-ranged temporal correlations. The characteristic features
of FLEFE motion is its validity beyond the (standard) Hurst exponent domain $0<
H\le1$ and the ability to explore properties of non-equilibrium systems---with
long-ranged correlations---that do not obey the FDT. The solution of the FLEFE
with a fractional derivative of order $\alpha>1/2$ and zero initial conditions
is given by the RL-FBM (FBM II) process with Hurst exponent $H=\alpha-1/2$. The
increments or the MSI of RL-FBM become stationary ($t$-independent) in the
domain $1/2<\alpha<3/2$ in the long-time limit; however, they remain
nonstationary when $\alpha\ge3/2$. We also showed that the $(n+1)$th MSI
restores stationarity for $(2n+1)/2<\alpha<(2n+3)/2$ in the long-time limit.
The nonequivalence of the TAMSD and MSD is observed in the whole regime $\alpha
>1/2$, leading to spurious non-ergodicity in the long time limit when $1/2<
\alpha <3/2$. More specifically, the mean TAMSD of RL-FBM converges to the MSI
rather than to the MSD. We also investigated the influence of the aging time
$t_a$ on the MSD, TAMSD, and ACF of the increments. In the limit of strong
aging, we demonstrated that ergodicity is restored, i.e., the disparity between
aged MSD and aged mean TAMSD vanish. Under strong aging, the increment-ACF
becomes stationary and solely depends on the lag time $\Delta$ in the regime
$1/2<\alpha<3/2$, while it depends on the aging time $t_a$ when $\alpha\geq3/2$.
This contrasts standard FBM and FLE which are both ergodic processes and
always independent of $t_a$. We thus promote the MSI and the $(n+1)$th order
MSI for data analysis, e.g., in terms of statistical observables
\cite{weigel2011,sikora2017,weiss2013,metzler2014,barkai2012}, machine learning
\cite{seckler2022,seckler2023,manzo2021,szwabinski2019,henrik,gorka2021,granik2019},
or Bayesian analyses \cite{thapa2018,kevin}.

In non-equilibrium bio-systems, as well as other complex systems, the FLEFE may
indeed be a more suitable candidate for evaluating the observed stochastic
dynamics than the FBM and FLE approaches. Generally speaking, FLEFE
motion will be relevant when we have non-equilibrium initial conditions, when
the measurement is started immediately after inserting tracer particles such
as colloidal beads or quantum dots into the medium, without waiting that they
equilibrate with the medium \cite{sabri}. One might also test nonequilibrium
effects by deliberately moving tracer particles out of equilibrium, e.g., by
puilling on them with optical tweezers \cite{lene}. Furthermore, the movement
of birds, animals, or bacteria is always in non-equilibrium, such that the use
of the FLEFE model appears preferable. This also holds true for inherently
nonequilibrium systems such as financial market data or climate dynamics.
We also note that RL-FBM (the solution of the FLEFE) can be
easily modified by allowing the exponent to vary over time or space, as it
is defined from a non-equilibrated initial condition at $t=0$. In contrast,
the FBM defined by Mandelbrot is less adaptable due to its definition in an
equilibrated state starting from $t=-\infty$ \cite{mandelbrot1968}. Moreover,
recent single-particle-tracking experiments revealed that intracellular
transport of tracers of various sizes in cells is often not only anomalous,
but also heterogeneous in time and space \cite{korabel2022, waigh2023,
korabel2021, han2020, worlitzer2022, shaban2020}. This implies that a single
diffusion exponent and diffusing coefficient are insufficient to describe the
underlying physical phenomena. In this sense, it is of interest to consider
the FLEFE motion when both diffusivity and diffusion exponent are functions
of time \cite{wang2023, slezak2023} or space or even randomly chosen from
certain distribution \cite{balcerek2022}. In fact, RL-FBM with time-dependent
exponents has been successfully validated in terms of the MSD and the power
spectral density through recent experimental measurements of quantum dot
motion within the cytoplasm of live mammalian cells, as observed by
single-particle tracking \cite{balcerek2023}. We hope that our development
of the RL-FBM model, especially in the context ergodicity, will offer
experimentalists new insights into understanding the role of heterogeneity
in anomalous diffusion transport phenomena.

One more relevant application of the FLEFE model may be in the physics of
climate. The stochastic Markovian energy balance models (EBM) introduced by
Hasselmann \cite{hasselmann1976} and recognized by the award of the 2021 Nobel
Prize for Physics, has achieved a great success in climate modeling. Due to
the complex intake and dissipation of energy when the FDT may be broken,
and taking memory effects of the global temperature anomaly into account,
the FLEFE and its generalizations may be more adequate tools to describe
the climate dynamics in the long time limit, see e.g., the recent review
\cite{watkins2024} and references therein.

In the future, other potential issues for the FLEFE process are to be
considered, including  harmonic potential, reflecting or absorbing
boundaries, resetting search \cite{tal2020} and delay effects
\cite{giuggioli2016}.

\begin{acknowledgments}

We acknowledge support from the German Science Foundation (DFG grant ME
1535/16-1), NSF-BMBF CRCNS (Grant No. 2112862/STAXS), and the National
Natural Science Foundation of China (Grant No. 12171466).

\end{acknowledgments}

\appendix

\section{MSI}
\label{AppendixA}

Considering the MSI given by Eq.~(\ref{ST}) we find
\begin{eqnarray}\label{a1}
\nonumber
\left< x^2_{\Delta}(t)\right>&=&\left<[x(t+\Delta)-x(t)]^2\right>\\
&=&\frac{2K_{\alpha}t^{2\alpha-1}}{(2\alpha-1)\Gamma(\alpha)^2}\left[\left(1+
\frac{\Delta}{t}\right)^{2\alpha-1}+1\right]-\frac{4K_{\alpha}t^{\alpha}(t+
\Delta)^{\alpha-1}}{\alpha\Gamma(\alpha)^2}\times{}_2F_1(1-\alpha,1;1+\alpha;
\frac{t}{t+\Delta}).
\end{eqnarray}
Using the Pfaff transformation (Eq.~(15.3.4) in  Ref.~\cite{abramowitz1964})
for the hypergeometric function in Eq.~(\ref{a1}), we obtain
\begin{equation}
\label{a2}
\frac{4K_{\alpha}t^{\alpha}(t+\Delta)^{\alpha-1}}{\alpha\Gamma(\alpha)^2}
\times{}_2F_1(1-\alpha,1;1+\alpha;\frac{t}{t+\Delta})=\frac{4K_{\alpha}t^{\alpha}
\Delta^{\alpha-1}}{\alpha\Gamma(\alpha)^{2}}\times{}_2F_1(1-\alpha,\alpha;1+
\alpha;-\frac{t}{\Delta}).
\end{equation}
With the relation between the hypergeometric function and the $H$-function,
Eq.~(1.131) in  Ref.~\cite{mathai2010} as well as Eq.~(8.3.2.7) in
Ref.~\cite{prudnikov1990}, we get
\begin{equation}
\label{a3}
_2F_1\left(1-\alpha,\alpha;\alpha+1;-\frac{t}{\Delta}\right)=\frac{\alpha}{
\Gamma(1-\alpha)}H_{2,2}^{1,2}\left[\frac{t}{\Delta}\left|\begin{array}{l}
(\alpha,1),(1-\alpha,1)\\(0,1),(-\alpha,1)\end{array}\right.\right]=\frac{
\alpha}{\Gamma(1-\alpha)}H_{2,2}^{2,1}\left[\frac{\Delta}{t}\left|
\begin{array}{l}(1,1),(1+\alpha,1)\\(1-\alpha,1),(\alpha,1)\end{array}\right.
\right].
\end{equation}
This yields an alternative form for the MSI, namely,
\begin{equation}
\label{a4}
\left< x^2_{\Delta}(t)\right>=\frac{2K_{\alpha}t^{2\alpha-1}}{(2\alpha-1)
\Gamma(\alpha)^2}\left[\left(1+\frac{\Delta}{t}\right)^{2\alpha-1}+1\right]
-\frac{4K_{\alpha}t^{\alpha}\Delta^{\alpha-1}}{\Gamma(\alpha)^2\Gamma(1-\alpha)}
H_{2,2}^{2,1}\left[\frac{\Delta}{t}\left|\begin{array}{l}(1,1),(1+\alpha,1)\\
(1-\alpha,1),(\alpha,1)\end{array}\right.\right].
\end{equation}
The first term in Eq.~(\ref{a4}) can be transformed into the series
\begin{eqnarray}
\label{a5}
\nonumber
\mathrm{first~term}&=&\frac{2K_{\alpha}t^{2\alpha-1}}{(2\alpha-1)\Gamma(\alpha)^2}
\left[2+\sum_{k=1}^{+\infty}\binom{2\alpha-1}{k}\left(\frac{\Delta}{t}\right)
^k\right]\\
&=&\frac{2K_{\alpha}t^{2\alpha-1}}{\Gamma(\alpha)^2}\left[\frac{2}{2\alpha-1}+
\frac{\Delta}{t}+(\alpha-1)\left(\frac{\Delta}{t}\right)^2+O\left(\left(\frac{
\Delta}{t}\right)^3\right)\right].
\end{eqnarray}

To expand the second term (the $H$-function) in Eq.~(\ref{a4}), one needs to be
more careful since different expansions need to be implemented based on the
value of $\alpha$. 

When $\alpha\neq1,~3/2,~2,~ 5/2,...$, we are allowed to use expansion (8.2.3.3)
in Ref.~\cite{prudnikov1990} at $\Delta\ll t$, 
\begin{eqnarray}
\nonumber
H_{2,2}^{2,1}\left[\frac{\Delta}{t}\left|\begin{array}{l}(1,1),(1+\alpha,1)\\(1
-\alpha,1),(\alpha,1)\end{array}\right.\right]&=&\sum_{k=0}^{+\infty}\frac{(-1)^k
\Gamma(1+k-\alpha)}{(2\alpha-1-k)k!}\left(\frac{\Delta}{t}\right)^{1-\alpha+k}\\
&&+ \sum_{k=0}^{+\infty}\frac{(-1)^k\Gamma(1-k-2\alpha)\Gamma(k+\alpha)}{\Gamma(1
-k)k!}\left(\frac{\Delta}{t}\right)^{\alpha+k}.
\label{a6}
\end{eqnarray}
Since the Gamma function has simple poles at non-positive integers \cite{olver2010},
only the single term $k=0$ survives in the second summation of Eq.~(\ref{a6}). Then
the second term in Eq.~(\ref{a4}) can be written as
\begin{eqnarray}
\label{a7}
\nonumber
\mathrm{second~term}&=&\frac{4K_{\alpha}t^{2\alpha-1}}{\Gamma(\alpha)^2\Gamma
(1-\alpha)}\left[\sum_{k=0}^{+\infty}\frac{(-1)^k\Gamma(1+k-\alpha)}{(2\alpha
-1-k)k!}\left(\frac{\Delta}{t}\right)^{k}+\frac{2\Gamma(1-2\alpha)\Gamma(\alpha)}
{\Gamma(1-\alpha)}\left(\frac{\Delta}{t}\right)^{2\alpha-1}\right]\\
\nonumber
&=&\frac{2K_{\alpha}t^{2\alpha-1}}{\Gamma(\alpha)^{2}}\Bigg[\frac{2}{2\alpha-1}
+\frac{\Delta}{t}+\frac{(2-\alpha)(1-\alpha)}{2\alpha-3}\left(\frac{\Delta}
{t}\right)^2+O\left(\left(\frac{\Delta}{t}\right)^3\right)\\
&&+\frac{2\Gamma(1-2\alpha)\Gamma(\alpha)}{\Gamma(1-\alpha)}\left(\frac{\Delta}{t}
\right)^{2\alpha-1}\Bigg].
\end{eqnarray}  

Substituting Eqs.~(\ref{a5}) and (\ref{a7}) into Eq.~(\ref{a4}), one can see
that the main order terms proportional to $t^{2\alpha-1}\left(\Delta/t\right)^0$
and $t^{2\alpha-1}\left(\Delta/t\right)^1$ cancel out. With the main remaining
terms we obtain
\begin{eqnarray}
\label{a8}
\langle x^2_{\Delta}(t)\rangle&=&\frac{2K_{\alpha}t^{2\alpha-1}}{\Gamma(
\alpha)^2}\Bigg[\frac{(\alpha-1)^2}{2\alpha-3}\left(\frac{\Delta}{t}\right)^2
-\frac{2\Gamma(1-2\alpha)\Gamma(\alpha)}{\Gamma(1-\alpha)}\left(\frac{
\Delta}{t}\right)^{2\alpha-1}\Bigg].
\end{eqnarray}
When $1/2<\alpha<3/2$, the MSI at long times $t\gg\Delta$ in Eq.~(\ref{a8}) is
dominated by the second term; using Euler's reflection formula of the Gamma
function, $\Gamma(z)\Gamma(1-z)=\pi/\sin(\pi z)$, for $z\notin\mathbb{Z}$, we
arrive at the approximation
\begin{equation}
\label{a9}
\langle x^2_{\Delta}(t)\rangle\sim-\frac{4K_{\alpha}\Gamma(1-2\alpha)}{\Gamma(
\alpha)\Gamma(1-\alpha)}\Delta^{2 \alpha-1}=\frac{2K_{\alpha}}{\Gamma(2\alpha)
|\cos(\pi\alpha)|}\Delta^{2\alpha-1}.
\end{equation}
When $\alpha>3/2$, the MSI in Eq.~(\ref{a8}) is dominated by the first term and
asymptotically reads
\begin{equation}
\label{a10}
\langle x^2_{\Delta}(t)\rangle\sim\frac{2K_{\alpha}(\alpha-1)^2}{(2\alpha-3)
\Gamma(\alpha)^2}t^{2\alpha-3}\Delta^2.   
\end{equation}
We note that for $\alpha=1,~3/2,~2,~ 5/2,\ldots$ are all the singular points in
the series expansion (\ref{a6}) that need to be excluded. Next, we consider the
cases $\alpha=3/2,~ 5/2,~7/2,\ldots,$ and $\alpha=~1,~2,~3,\ldots,$ respectively.
In the former situation, the series of the $H$-function should be more complicated,
with power-logarithmic expansions; and in the latter, the $H$-function expansion
fails. In that case we expand the hypergeometric function in Eq.~(\ref{a1}) rather
than the $H$-function in Eq.~(\ref{a4}).

When $\alpha=3/2,~ 5/2,~7/2,\ldots,$, using the power-logarithmic expansions
(1.4.7) from Ref.~\cite{kilbas1999} in Eq.~(\ref{a4}), we find
\begin{eqnarray}
\label{a11}
\nonumber
H_{2,2}^{2,1}\left[\frac{\Delta}{t}\left|\begin{array}{l}(1,1),(1+\alpha,1)\\
(1-\alpha,1),(\alpha,1)\end{array}\right.\right]&=&\sum_{k=0}^{2\alpha-2}\frac{
(-1)^k\Gamma(k+1-\alpha)}{(2\alpha-k-1)k!}\left(\frac{\Delta}{t}\right)^{1-\alpha
+k}-\frac{\Gamma(\alpha)}{\Gamma(2\alpha)}\left(\frac{\Delta}{t}\right)^{\alpha}
\ln\left(\frac{\Delta}{t}\right)\\
&&+O\left(\left(\frac{\Delta}{t}\right)^{\alpha}\ln \left(\frac{\Delta}{t}
\right)\right).
\end{eqnarray}
Then the second term in the MSI (\ref{a4}) asymptotically reads for $t\gg\Delta$
\begin{eqnarray}
\label{a12}
\nonumber
\mathrm{second~term}
&\sim&\frac{4K_{\alpha}t^{2\alpha-1}}{\Gamma(\alpha)^{2}\Gamma(1-\alpha)}\Bigg[
\sum_{k=0}^{2\alpha-2}\frac{(-1)^{k}\Gamma(k+1-\alpha)}{(2\alpha-k-1)k!}\left(
\frac{\Delta}{t}\right)^k-\frac{\Gamma(\alpha)}{\Gamma(2\alpha)}\left(\frac{
\Delta}{t}\right)^{2\alpha-1}\ln\left(\frac{\Delta}{t}\right)\\
&&+O\left(\left(\frac{\Delta}{t}\right)^{2\alpha-1}\ln \left(\frac{\Delta}{t}
\right)\right)\Bigg].
\end{eqnarray}
When $\alpha=3/2$, this second term is
\begin{eqnarray}
\label{a13}
\mathrm{second~term}
&\sim&\frac{8 K_{3/2}t^2}{\pi}\Bigg[1+\frac{\Delta}{t}-\frac{1}{4}\left(
\frac{\Delta}{t}\right)^2\ln\left(\frac{\Delta}{t}\right)+O\left(\left(\frac{
\Delta}{t}\right)^2\ln\left(\frac{\Delta}{t}\right)\right)\Bigg].
\end{eqnarray}
Substituting Eqs.~(\ref{a13}) and (\ref{a5}) into Eq.~(\ref{a4}), we arrive at
the MSI with $\alpha=3/2$ at long times $t\gg\Delta$,
\begin{eqnarray}
\label{a14}
\langle x^2_{\Delta}(t)\rangle&\sim&\frac{2K_{3/2}}{\pi}\Delta^2\ln\left(
\frac{t}{\Delta}\right).
\end{eqnarray}
When $\alpha\geq5/2$, namely, for $\alpha=5/2,~7/2,~9/2,\ldots$, the second term
asymptotically reads
\begin{eqnarray}
\label{a15}
\nonumber
\mathrm{second~term} & \sim& \frac{2 K_\alpha t^{2\alpha-1}}{\Gamma(\alpha)^2}\Bigg[\frac{2}{2 \alpha-1}+\frac{\Delta}{t}+\frac{(2-\alpha)(1-\alpha)}{(2 \alpha-3)}\left(\frac{\Delta}{t}\right)^{2}\\
&&+\frac{2\Gamma(\alpha)}{\Gamma(2\alpha)\Gamma(1-\alpha)}\left(\frac{\Delta}{t}\right)^{2\alpha-1} \ln \left(\frac{\Delta}{t}\right)+O\left(\left(\frac{\Delta}{t}\right)^{2\alpha-1}\ln \left(\frac{\Delta}{t}\right)\right)\Bigg]. 
\end{eqnarray}
Substituting Eq.~(\ref{a15}) and Eq.~(\ref{a5}) into Eq.~(\ref{a4}), we obtain the MSI with the leading term as
\begin{eqnarray}
\label{a16}
\nonumber
\langle x^2_{\Delta}(t)\rangle&\sim&\frac{2K_{\alpha}(\alpha-1)^2}{(2\alpha-3)
\Gamma(\alpha)^2}t^{2\alpha-3}\Delta^2\Bigg[1-\frac{2(2\alpha-3)\Gamma(\alpha)}{
\Gamma(2\alpha)\Gamma(1-\alpha)(\alpha-1)^2}\left(\frac{\Delta}{t}\right)^{2
\alpha-3}\ln\left(\frac{\Delta}{t}\right)\Bigg]\\
&\sim&\frac{2K_{\alpha}(\alpha-1)^2}{(2\alpha-3)\Gamma(\alpha)^2}t^{2\alpha-3}
\Delta^2.  
\end{eqnarray}
This result is identical to Eq.~(\ref{a10}). 

When $\alpha=1,~2,~3,~4,\ldots$, we turn to the hypergeometric function (\ref{a2})
and then use formulas (15.8.6) and (15.2.4) in Ref.~\cite{olver2010} with $\alpha
\in\mathbb{N}^+$,
\begin{equation}
\label{a17}
{}_2 F_1\left(1-\alpha,\alpha;1+\alpha;-\frac{t}{\Delta}\right)=\frac{(\alpha)
_{\alpha-1}}{(1+\alpha)_{\alpha-1}}\left(\frac{t}{\Delta}\right)^{\alpha-1}{}_2
F_1\left(1-\alpha,1-2\alpha;2-2\alpha;-\frac{\Delta}{t}\right)
\end{equation}
and
\begin{equation}
\label{a18}
{}_2F_1\left(1-\alpha,1-2\alpha; 2-2\alpha;-\frac{\Delta}{t}\right)=\sum_{n=0}
^{\alpha-1}\binom{\alpha-1}{n}\frac{(1-2\alpha)_n}{(2-2\alpha)_n}\left(\frac{
\Delta}{t}\right)^n,
\end{equation}
where $(q)_n$ is the (rising) Pochhammer symbol,
\begin{equation}
(q)_n=\left\{\begin{array}{ll}1,&n=0\\q(q+1)\cdots(q+n-1),&n>0\end{array}
\right..
\end{equation}
We then rewrite Eq.~(\ref{a2}) in the form with $t>\Delta$
\begin{eqnarray}
\nonumber
\frac{4K_\alpha t^\alpha\Delta^{\alpha-1}}{\alpha\Gamma(\alpha)^2}{}_2F_1\left(
1-\alpha,\alpha;1+\alpha;-\frac{t}{\Delta}\right)&=&\frac{4K_\alpha t^{2\alpha-
1}}{\Gamma(\alpha)^2}\Bigg[\sum_{n=0}^{\alpha-1}\frac{\binom{\alpha-1}{n}}{2
\alpha-1-n}\left(\frac{\Delta}{t}\right)^{n}\Bigg]\\
&&\hspace*{-2cm}=\frac{2K_\alpha t^{2\alpha-1}}{\Gamma(\alpha)^2} \Bigg[\frac{
2}{2\alpha-1}+\frac{\Delta}{t}+\frac{(\alpha-1)(\alpha-2)}{2\alpha-3}\left(
\frac{\Delta}{t}\right)^2+\ldots+\left(\frac{\Delta}{t}\right)^{\alpha-1}\Bigg].
\label{a19}
\end{eqnarray} 
Substituting Eq.~(\ref{a19}) into the MSI (\ref{a1}) and considering the
expansion of the first term in Eq.~(\ref{a5}), one can easily find that at
long times $t\gg\Delta$ the cases $\alpha=1$ (here only the constant remains
in Eq.~(\ref{a19})) and $\alpha=2,3,4,\ldots$ agree with the asymptotics
(\ref{a9}) and (\ref{a10}), respectively.

\section{TAMSD}
\label{AppendixB}

According to the definition (\ref{TAMSD}) of the TAMSD, the mean TAMSD of
RL-FBM can be derived as
\begin{eqnarray}
\nonumber
\left<\overline{\delta^2(\Delta)}\right>&=&\frac{1}{T-\Delta}\int_0^{T-\Delta}
\langle\left[x(t+\Delta)-x(t)\right]^2\rangle dt\\
&=&\frac{K_\alpha}{\alpha(2\alpha-1)\Gamma(\alpha)^2}\frac{T^{2\alpha}-\Delta^{
2\alpha}}{T-\Delta}+\frac{K_\alpha}{\alpha(2\alpha-1)\Gamma(\alpha)^2}(T-\Delta)
^{2\alpha-1}-I_{\alpha}(\Delta,T),
\label{b1}
\end{eqnarray}
where
\begin{equation}
\label{b2}
I_{\alpha}(\Delta,T)=\frac{4K_\alpha}{\alpha\Gamma(\alpha)^2}\frac{1}{T-\Delta}
\int_0^{T-\Delta}(t+\Delta)^{\alpha-1}t^\alpha\times{}_2F_1\left(1-\alpha,1;1+
\alpha;\frac{t}{t+\Delta}\right)dt. 
\end{equation}
Applying the same transformation from the hypergeometric function to the
$H$-function as in Eqs.~(\ref{a2}) and (\ref{a3}) and the identities
(1.16.4.1) and (8.3.2.1) for the  $H$-function in Ref.~\cite{prudnikov1990}, we
obtain
\begin{eqnarray}
\nonumber
I_{\alpha}(\Delta,T)&=&\frac{4K_\alpha}{\Gamma(\alpha)^2\Gamma(1-\alpha)}\frac{
\Delta^{2\alpha}}{(T-\Delta)}\int_0^{\frac{T-\Delta}{\Delta}}s^\alpha H_{2,2}^{
1,2}\left[s\left|\begin{array}{l}(\alpha,1),(1-\alpha,1)\\(0,1),(-\alpha,1)
\end{array}\right.\right]ds\\
\nonumber
&=&\frac{4K_\alpha}{\Gamma(\alpha)^2\Gamma(1-\alpha)}\frac{\Delta^{2\alpha}}{(T
-\Delta)}\frac{(T-\Delta)^{\alpha+1}}{\Delta^{\alpha+1}}H_{3,3}^{1,3}\left[\frac{
T-\Delta}{\Delta}\left|\begin{array}{l}(-\alpha,1),(\alpha,1),(1-\alpha,1)\\
(0,1),(-\alpha,1),(-1-\alpha,1)\end{array}\right.\right]\\
&=&\frac{4K_\alpha}{\Gamma(\alpha)^2\Gamma(1-\alpha)}\frac{(T-\Delta)^{\alpha} }{
\Delta^{1-\alpha}}H_{3,3}^{1,3}\left[\frac{T-\Delta}{\Delta}\left|\begin{array}{l}
(-\alpha,1),(\alpha,1),(1-\alpha,1)\\(0,1),(-1-\alpha,1),(-\alpha,1)\end{array}
\right.\right].
\end{eqnarray}
Moreover, using the identities (8.3.2.6) and (8.3.2.7) in Ref.~\cite{prudnikov1990}
to reduce to lower orders of the $H$-function, we find
\begin{eqnarray}
\nonumber
H_{3,3}^{1,3}\left[\frac{T-\Delta}{\Delta}\left|\begin{array}{l}(-\alpha,1),
(\alpha,1),(1-\alpha,1)\\(0,1),(-1-\alpha,1),(-\alpha,1)\end{array}\right.
\right]&=&H_{2,2}^{1,2}\left[\frac{T-\Delta}{\Delta}\left|\begin{array}{l}
(\alpha,1),(1-\alpha,1)\\(0,1),(-1-\alpha,1)\end{array}\right.\right]\\
&=&H_{2,2}^{2,1}\left[\frac{\Delta}{T-\Delta}\left|\begin{array}{l}(1,1),
(2+\alpha,1)\\(1-\alpha,1),(\alpha,1)\end{array}\right.\right],
\end{eqnarray}
such that the the third term in Eq.~(\ref{b1}) yields as
\begin{equation}
\label{b3}
I_{\alpha}(\Delta,T)=\frac{4K_\alpha}{\Gamma(\alpha)^2\Gamma(1-\alpha)}\frac{
(T-\Delta)^{\alpha}}{\Delta^{1-\alpha}}H_{2,2}^{2,1}\left[\frac{\Delta}{T-
\Delta}\left|\begin{array}{l}(1,1),(2+\alpha,1)\\(1-\alpha,1),(\alpha,1)
\end{array}\right.\right]. 
\end{equation}
Therefore the alternative expression for the mean TAMSD can be obtained
explicitly,
\begin{eqnarray}
\label{b4}
\nonumber
\left<\overline{\delta^2(\Delta)}\right>&=&\frac{K_\alpha}{\alpha(2
\alpha-1)\Gamma(\alpha)^2}\frac{T^{2\alpha}-\Delta^{2\alpha}}{T-\Delta}+
\frac{K_\alpha}{\alpha(2\alpha-1)\Gamma(\alpha)^2}(T-\Delta)^{2\alpha-1}\\
&&-\frac{4 K_\alpha}{\Gamma(\alpha)^2\Gamma(1-\alpha)}\frac{(T-\Delta)^{
\alpha}}{\Delta^{1-\alpha}}H_{2,2}^{2,1}\left[\frac{\Delta}{T-\Delta}\left|
\begin{array}{l}(1,1),(2+\alpha,1)\\(1-\alpha,1),(\alpha,1)\end{array}
\right.\right].
\end{eqnarray}

To obtain the approximations of the TAMSD at long times $T\gg\Delta$, we
consider the first and second terms in Eq.~(\ref{b4}),
\begin{eqnarray}
\label{b5}
\nonumber
\mathrm{first~term}&=&\frac{K_{\alpha}T^{2\alpha-1}}{\alpha(2\alpha-1)\Gamma(
\alpha)^2}\left[1-\left(\frac{\Delta}{T}\right)^{2\alpha}\right]\left(1-\frac{
\Delta}{T}\right)^{-1}\\
&=&\frac{K_{\alpha}T^{2\alpha-1}}{\Gamma(\alpha)^2}\left[\frac{1}{\alpha(2
\alpha-1)}+\frac{1}{\alpha(2\alpha-1)}\frac{\Delta}{T}+\frac{1}{\alpha(2\alpha
-1)}\left(\frac{\Delta}{T}\right)^2+O\left(\left(\frac{\Delta}{T}\right)^2
\right)\right],
\end{eqnarray}
and
\begin{eqnarray}
\nonumber
\mathrm{second~term}&=&\frac{K_{\alpha}}{\alpha(2\alpha-1)\Gamma(\alpha)^2}
(T-\Delta)^{2 \alpha-1}\\
&=&\frac{K_{\alpha}T^{2\alpha-1}}{\Gamma(\alpha)^2}\left[\frac{1}{\alpha(2
\alpha-1)}-\frac{\Delta}{\alpha T}+\frac{\alpha-1}{\alpha}\left(\frac{
\Delta}{T}\right)^2+O\left(\left(\frac{\Delta}{T}\right)^2\right)\right].
\label{b6}
\end{eqnarray}

The approach to tackle the third term in Eq.~(\ref{b4}) is analogous to that
for the MSI. When $\alpha\neq1,~3/2,~2,~5/2,\ldots$, using formula (8.2.3.3)
from Ref.~\cite{prudnikov1990}, we get 
\begin{eqnarray}
\nonumber
H_{2,2}^{2,1}\left[\frac{\Delta}{T-\Delta}\left|\begin{array}{l}(1,1),(2+
\alpha,1)\\(1-\alpha,1),(\alpha,1)\end{array}\right.\right]&=&\sum_{k=0}^{
\infty}\frac{(-1)^k\Gamma(k+1-\alpha)}{(2\alpha-k)(2\alpha-k-1)k!}\left(
\frac{\Delta}{T-\Delta}\right)^{1-\alpha+k}\\
&&+\sum_{k=0}^{\infty}\frac{(-1)^k\Gamma(1-2\alpha-k)\Gamma(\alpha+k)}{\Gamma(
2-k)k!}\left(\frac{\Delta}{T-\Delta}\right)^{\alpha+k}.
\label{b7}
\end{eqnarray}

In Eq.~(\ref{b7}) only two terms remain in the second series. With Eq.~(\ref{b7})
we obtain the third term of Eq.~(\ref{b4}) at long times $T\gg\Delta$, 
\begin{eqnarray}
\nonumber
I_{\alpha}(\Delta,T)&=&\frac{4K_\alpha T^{2\alpha-1}}{\Gamma(\alpha)^2\Gamma(
1-\alpha)}\sum_{k=0}^{\infty}\frac{\Gamma(k+1-\alpha)(-1)^{k}}{(2\alpha-k)(2
\alpha-k-1)k!}\left(\frac{\Delta}{T}\right)^{k}\left(1-\frac{\Delta}{T}
\right)^{2\alpha-1-k}\\
\nonumber
&&+\frac{4K_\alpha\Delta^{2\alpha-1}}{\Gamma(\alpha)^{2}\Gamma(1-\alpha)}\sum_{k=0}
^1\frac{(-1)^k\Gamma(1-2\alpha-k)\Gamma(\alpha+k)}{\Gamma(2-k)k!}\left(\frac{
\Delta}{T-\Delta}\right)^k\\
\nonumber
&\sim&\frac{2K_\alpha T^{2\alpha-1}}{\Gamma(\alpha)^2}\Bigg[\frac{1}{\alpha(2
\alpha-1)}+\frac{1-\alpha}{\alpha(2\alpha-1)}\frac{\Delta}{T}+\left(\frac{1-
\alpha}{\alpha(2\alpha-1)}+\frac{\alpha-2}{2(2\alpha-3)}\right)\left(
\frac{\Delta}{T}\right)^2+O\left(\left(\frac{\Delta}{T}\right)^2\right)\Bigg]\\
&&+\frac{4K_\alpha\Gamma(1-2\alpha)T^{2\alpha-1}}{\Gamma(\alpha)\Gamma(1-\alpha)}
\Bigg[\left(\frac{\Delta}{T}\right)^{2\alpha-1}+\frac{1}{2}\left(\frac{\Delta}{
T}\right)^{2\alpha}
\Bigg]. 
\label{b8}
\end{eqnarray}
Substituting Eqs.~(\ref{b5}), (\ref{b6}), and (\ref{b8}) into Eq.~(\ref{b4}), one
can see that the main order terms proportional to $T^{2\alpha-1}\left(\frac{
\Delta}{T}\right)^0$ and $T^{2\alpha-1}\left(\frac{\Delta}{T}\right)^1$ cancel
out. We thus obtain the mean TAMSD of RL-FBM at long times $T\gg\Delta$, 
\begin{equation}
\label{TAMSDcase1} 
\left<\overline{\delta^2(\Delta)}\right>\sim\frac{2K_{\alpha}T^{2\alpha-1}}{
\Gamma(2\alpha)|\cos(\pi\alpha)|}\Bigg[\left(\frac{\Delta}{T}\right)^{2\alpha
-1}+\frac{1}{2}\left(\frac{\Delta}{T}\right)^{2\alpha}\Bigg]+\frac{(\alpha-
1)K_{\alpha}T^{2\alpha-1}}{(2\alpha-3)\Gamma(\alpha)^2}\left(\frac{
\Delta}{T}\right)^2. 
\end{equation}
Moreover, for $1/2<\alpha<3/2$ the mean TMASD is dominated by the first term,
\begin{equation}
\label{b10}
\left<\overline{\delta^2(\Delta)}\right>\sim\frac{2K_{\alpha}}{\Gamma(2\alpha)
|\cos(\pi\alpha)|}\Delta^{2\alpha-1}, 
\end{equation}
whereas for $\alpha>3/2$, the third term is dominant,
\begin{equation}
\label{b11} 
\left<\overline{\delta^2(\Delta)}\right>\sim\frac{(\alpha-1)K_{\alpha}}{(2
\alpha-3)\Gamma(\alpha)^2}\Delta^{2}T^{2\alpha-3}.
\end{equation}

Conversely, for $\alpha=3/2,~5/2,~7/2,\ldots$ the power series expansion (\ref{b7})
fails in Eq.~(\ref{b4}). Instead, we use the power-logarithmic expansion (1.4.7)
in Ref.~\cite{kilbas1999},
\begin{eqnarray}
\label{b12}
\nonumber
H_{2,2}^{2,1}\left[\frac{\Delta}{T-\Delta}\left|\begin{array}{l}(1,1),(2+\alpha,
1)\\(1-\alpha,1),(\alpha,1)\end{array}\right.\right]&=&\sum_{k=0}^{2\alpha-2}
\frac{(-1)^k\Gamma(k+1-\alpha)}{(2\alpha-k)(2\alpha-k-1)k!}\left(\frac{\Delta}{
T-\Delta}\right)^{1-\alpha+k}\\
&&-\frac{\Gamma(\alpha)}{\Gamma(2\alpha)}\left(\frac{\Delta}{T-\Delta}\right)
^{\alpha}\ln\left(\frac{\Delta}{T-\Delta}\right)+O\left(\left(\frac{\Delta}{T
-\Delta}\right)^{\alpha}\ln\left(\frac{\Delta}{T-\Delta}\right)\right),
\end{eqnarray}
and then the third term (\ref{b3}) in the mean TAMSD (\ref{b4}) in the long
time limit $T\gg\Delta$ reads
\begin{eqnarray}
\nonumber
I_{\alpha}(\Delta,T)&=&\frac{4K_\alpha T^{2\alpha-1}}{\Gamma(\alpha)^2\Gamma(1-
\alpha)}\Bigg[\sum_{k=0}^{2\alpha-2}\frac{\Gamma(k+1-\alpha)(-1)^k}{(2\alpha-k)
(2\alpha-k-1)k!}\left(\frac{\Delta}{T}\right)^k\left(1-\frac{\Delta}{T}\right)
^{2\alpha-1-k}-\frac{\Gamma(\alpha)}{\Gamma(2\alpha)}\left(\frac{\Delta}{T}
\right)^{2\alpha-1}\ln\left(\frac{\Delta}{T}\right)\\
&&+O\left(\left(\frac{\Delta}{T}\right)^{2\alpha-1}\ln\left(\frac{\Delta}{T}
\right)\right)\Bigg].
\label{b13}
\end{eqnarray}
In particular, when $\alpha=3/2$,
\begin{equation}
\label{b14}
I_{3/2}(\Delta,T)=\frac{8K_{3/2}T^2}{\pi}\Bigg[\frac{1}{3}-\frac{\Delta}{6T}
-\frac{1}{6}\left(\frac{\Delta}{T}\right)^2+\frac{1}{4}\left(\frac{\Delta}{T}
\right)^2\ln\left(\frac{\Delta}{T}\right)+O\left(\left(\frac{\Delta}{T}\right)^2
\ln\left(\frac{\Delta}{T}\right)\right)\Bigg].
\end{equation}
Substituting Eqs.~(\ref{b14}), (\ref{b5}), and (\ref{b6}) into Eq.~(\ref{b4}),
we obtain the mean TAMSD with the leading term in the long time limit $T\gg
\Delta$,
\begin{equation}
\label{b15}
\left<\overline{\delta^2(\Delta)}\right>\sim\frac{2K_{3/2}}{\pi}\Delta^2
\ln\left(\frac{T}{\Delta}\right).
\end{equation}
When $\alpha=5/2,~7/2,~9/2,\ldots$, Eq.~(\ref{b13}) reads
\begin{eqnarray}
\nonumber
I_{\alpha}(\Delta,T)&\sim&\frac{2K_\alpha T^{2\alpha-1}}{\Gamma(\alpha)^2}\Bigg[
\frac{1}{\alpha(2\alpha-1)}\left(1-\frac{\Delta}{T}\right)^{2\alpha-1}+\frac{1}{
(2\alpha-1)}\left(\frac{\Delta}{T}\right)\left(1-\frac{\Delta}{T}\right)^{2
\alpha-2}+\frac{\alpha-2}{2(2\alpha-3)}\left(\frac{\Delta}{T}\right)^{2}\left(
1-\frac{\Delta}{T}\right)^{2\alpha-3}\\
\nonumber
&&-\frac{2\Gamma(\alpha)}{\Gamma(2\alpha)\Gamma(1-\alpha)}\left(\frac{\Delta}{T}
\right)^{2\alpha-1}\ln\left(\frac{\Delta}{T}\right)+O\left(\left(\frac{
\Delta}{T}\right)^{2\alpha-1}\ln\left(\frac{\Delta}{T}\right)\right)\Bigg]\\
\nonumber
&\sim&\frac{2K_\alpha T^{2\alpha-1}}{\Gamma(\alpha)^2}\Bigg[\frac{1}{\alpha(2
\alpha-1)}+\frac{1-\alpha}{\alpha(2\alpha-1)}\left(\frac{\Delta}{T}\right)+
\left(\frac{1-\alpha}{\alpha(2\alpha-1)}+\frac{\alpha-2}{2(2 \alpha-3)}\right)
\left(\frac{\Delta}{T}\right)^2+O\left(\left(\frac{\Delta}{T}\right)^3\right)
\Bigg]\\
&&-\frac{4K_{\alpha}\Delta^{2\alpha-1}}{\Gamma(\alpha)\Gamma(2\alpha)\Gamma(1
-\alpha)}\ln\left(\frac{\Delta}{T}\right). 
\label{b16}
\end{eqnarray}
Substituting Eqs.~(\ref{b16}), (\ref{b5}), and (\ref{b6}) into Eq.~(\ref{b4}),
one can see that the main order terms proportional to $T^{2\alpha-1}\left(
\Delta/T\right)^0$ and $T^{2\alpha-1}\left(\Delta/T\right)^1$ cancel out. Thus,
we obtain the mean TAMSD of RL-FBM in the long time limit $T\gg\Delta$ as
\begin{eqnarray}
\label{b17}
\nonumber
\left<\overline{\delta^2(\Delta)}\right>&\sim& \frac{2K_{\alpha}}{\Gamma(
\alpha)\Gamma(1-\alpha)}\Delta^{2\alpha-1}\ln\left(\frac{\Delta}{T}\right)+
\frac{(\alpha-1)K_{\alpha}}{(2\alpha-3)\Gamma(\alpha)^2}\Delta^{2}T^{2\alpha
-3}\\
&\sim&\frac{(\alpha-1)K_{\alpha}}{(2\alpha-3)\Gamma(\alpha)^2}\Delta^2T^{
2\alpha-3}. 
\end{eqnarray}

For $\alpha=1,~2,~3,\ldots$ we make the variable transformation $z=t/(t+\Delta)$
in Eq.~(\ref{b2}) and use the integral formula (1. 153.11) for the hypergeometric
function in Ref.~\cite{prudnikov1990}. As result we have 
\begin{equation}
\label{b18}
I_{\alpha}(\Delta,T)=\frac{4K_{\alpha}}{\alpha(\alpha+1)\Gamma(\alpha)^2}T^{2
\alpha-1}\left(1-\frac{\Delta}{T}\right)^{\alpha}{}_2F_1\left(1-\alpha,2;
\alpha+2;1-\frac{\Delta}{T}\right).
\end{equation}
Again, considering the series expansion of the hypergeometric function (\ref{a18})
with integer $\alpha$, we arrive at the third term in the mean TAMSD (\ref{b1}),
\begin{eqnarray}
\label{b19}
I_{\alpha}(\Delta,T)&=&\frac{4K_{\alpha}T^{2\alpha-1}}{\alpha(\alpha+1)\Gamma(
\alpha)^2}\sum_{n=0}^{\alpha-1}(-1)^n\binom{\alpha-1}{n}\frac{(2)_n}{(2+\alpha)_n}
\left(1-\frac{\Delta}{T}\right)^{\alpha+n}.
\end{eqnarray}
In particular, for $\alpha=1$, 
\begin{equation}
\label{b20}
I_1(\Delta,T)=2K_1T\left(1-\frac{\Delta}{T}\right),    
\end{equation}
and for $\alpha\ge2$, 
\begin{eqnarray}
\label{b21}
I_{\alpha}(\Delta,T)&=&\frac{4K_{\alpha}T^{2\alpha-1}}{\alpha(\alpha+1)\Gamma(
\alpha)^2}\Bigg[A_0-A_1\frac{\Delta}{T}+A_2\left(\frac{\Delta}{T}\right)^2+
\ldots+(-1)^{\alpha-1}A_{\alpha-1}\left(\frac{\Delta}{T}\right)^{\alpha-1}\Bigg],
\end{eqnarray} 
where 
\begin{equation}
A_n=\sum_{k=0}^n\binom{\alpha}{n-k}\frac{{}_2F^{(k)}_1\left(1-\alpha,2;\alpha+2;
1\right)}{k!}.
\end{equation}
These factors can be obtained from the properties of the hypergeometric function
and its $k$th ($k\leq\alpha-1$) order derivative with integer $\alpha$ in the form
\begin{equation}
\label{b22}
{}_2F^{(k)}_1(1-\alpha,2;\alpha+2;1)=\frac{(1-\alpha)_k(2)_k}{(\alpha+2)_k}{}_2F
_1(1-\alpha+k,2+k;\alpha+2+k;1).
\end{equation} 
Here,
\begin{equation}
\label{b23}
{}_2F_1(1-\alpha+k,2+k;\alpha+2+k;1)=\frac{(\alpha)_{\alpha-1-k}}{(\alpha+2+k)_{
\alpha-1-k}}.  
\end{equation} 
Thus Eq.~(\ref{b21}) reads
\begin{eqnarray}
\label{b24}
\hspace*{-0.6cm}
I_{\alpha}(\Delta,T)&=&\frac{2K_{\alpha}T^{2\alpha-1}}{\Gamma(\alpha)^2}\Bigg[
\frac{1}{\alpha(2\alpha-1)}-\frac{\alpha-1}{\alpha(2\alpha-1)}\frac{\Delta}{T}
+\left(\frac{1}{2\alpha-1}-\frac{1}{\alpha}+\frac{\alpha-2}{2(2\alpha-3)}\right)
\left(\frac{\Delta}{T}\right)^2+O\left(\left(\frac{\Delta}{T}\right)^2\right)
\Bigg].
\end{eqnarray}
Substituting Eqs.~(\ref{b20}) and (\ref{b24}) into the mean TAMSD (\ref{b1}),respectively, and considering the expansions of the first and second terms,
Eqs.~(\ref{b5}) and (\ref{b6}), respectively, one can easily find that at long
times $t\gg\Delta$ the cases $\alpha=1$ and $\alpha\ge 2$ are in agreement with
the asymptotics in Eqs.~(\ref{b10}) and (\ref{b11}), respectively.

\section{Simulation approach for RL-FBM}
\label{AppendixC}

Here we provide our numerical approach to discretize the stochastic integral
representation given by Eq.~(\ref{solution}) to generate the trajectories of
RL-FBM $x(t)$ at discrete times $t_n=n\times\delta t$, $n=N_+$, for all
admissible values $\alpha>1/2$. To begin, we discretize the stochastic integral,
\begin{eqnarray}
\label{c1}
x(t_n)=\frac{\sqrt{2K_\alpha}}{\Gamma(\alpha)}\sum_{i=0}^{n-1}\int_{t_i}^{t_{
i+1}}\left(t_n-s\right)^{\alpha-1}\xi(s)ds.
\end{eqnarray}
The white Gaussian noise can be approximated by
\begin{eqnarray}
\label{c2}
\xi(s)=\nu_i/\sqrt{\delta t},
\end{eqnarray}
where $\nu_i$ is a normally distributed random variable with zero mean and unit
variance; then we have
\begin{eqnarray}
\nonumber
x(t_n)&=&\frac{\sqrt{2K_\alpha}}{\Gamma(\alpha)}\sum_{i=0}^{n-1}\left(\frac{
\nu_i}{\sqrt{\delta t}}\right)\int_{t_i}^{t_{i+1}}\left(t_n-s\right)^{\alpha
-1}ds\\
\nonumber
&=&\frac{\sqrt{2K_\alpha}}{\Gamma(\alpha)}\sum_{i=0}^{n-1}\left(\frac{\nu_i}{
\sqrt{\delta t}}\right)\frac{\left(t_n-t_i\right)^\alpha-\left(t_n-t_i-\delta
t\right)^\alpha}{\alpha}\\
&=&\frac{\sqrt{2K_\alpha}}{\Gamma(\alpha)}\sum_{i=0}^{n-1}\left(\frac{\nu_i}{
\sqrt{\delta t}}\right)w(t_n-t_i)\delta t,
\label{c3}
\end{eqnarray}
with the weight function
\begin{equation}
\label{c4}
w(t_n-s)=\frac{(t_n-s)^\alpha-(t_n-s-\delta t)^\alpha}{\alpha\delta t}.
\end{equation}
This weight function in Eq.~(\ref{c4}) is the approximation of the kernel $(t_n-s)^{\alpha-1}$ in Eq.~(\ref{c1}) and does not contain any singularity for all $\alpha>1/2$.

\section{Aged ACF of the increments}
\label{AppendixD}

The aged ACF of the increments $x^\delta(t)=x(t+\delta)-x(t)$ of RL-FBM is defined
as 
\begin{equation}
C(t_a,\Delta)=\langle x^\delta(t_a)x^\delta(t_a+\Delta)\rangle,
\end{equation}
where $\delta$ is a small time step obeying $\delta\ll\Delta$. When $\Delta=0$,
the ACF corresponds to the MSI or aged MSD. 

With this notation we obtain
\begin{eqnarray}
\nonumber
C(t_a,\Delta)&=&\frac{2K_{\alpha}}{\Gamma(\alpha)^2}\int_0^{t_a+\delta}(t_a+
\delta-u)^{\alpha-1}\left[(t_a+\Delta+\delta-u)^{\alpha-1}-(t_a+\Delta-u)^{
\alpha-1}\right]du\\
&&+\frac{2K_{\alpha}}{\Gamma(\alpha)^2}\int_0^{t_a}(t_a-u)^{\alpha-1}\left[
(t_a+\Delta-u)^{\alpha-1}-(t_a+\Delta+\delta-u)^{\alpha-1}\right]du.
\label{d1}
\end{eqnarray}
After the change of variables $q=t_a+\delta-u$ and $q=t_a-u$ in the first and
second integral, respectively, we find
\begin{eqnarray}
\nonumber
C(t_a,\Delta)&=&\frac{2K_{\alpha}}{\Gamma(\alpha)^2}\int_0^{t_a}q^{\alpha-1}(q
+\Delta)^{\alpha-1}\left[2-\left(1+\frac{\delta}{q+\Delta}\right)^{\alpha-1}
-\left(1-\frac{\delta}{q+\Delta}\right)^{\alpha-1}\right]dq\\
&&+\frac{2K_{\alpha}}{\Gamma(\alpha)^2}\int_{t_a}^{t_a+\delta}q^{\alpha-1}(q+
\Delta)^{\alpha-1}\left[1-\left(1-\frac{\delta}{q+\Delta}\right)^{\alpha-1}
\right]dq.
\label{d2}
\end{eqnarray}
For a small time step $\delta\to0$ we consider the series expansion under the
condition that $\Delta\gg\delta$ and neglect the higher order terms,
\begin{equation}
\label{d3}
\left(1\pm\frac{\delta}{q+\Delta}\right)^{\alpha-1}\sim1\pm(\alpha-1)\frac{
\delta}{q+\Delta}+\frac{(\alpha-1)(\alpha-2)}{2}\left(\frac{\delta}{q+\Delta}
\right)^2.
\end{equation}
We thus arrive at the ACF
\begin{eqnarray}
\nonumber
C(t_a,\Delta)&\sim&\frac{2K_{\alpha}}{\Gamma(\alpha)^2}\left[(\alpha-1)(2-\alpha)
\delta^2\int_0^{t_a}q^{\alpha-1}(q+\Delta)^{\alpha-3}dq+(\alpha-1)\delta\int_{t_a}
^{t_a+\delta} q^{\alpha-1}(q+\Delta)^{\alpha-2}dq\right]\\
&=&\frac{2K_\alpha(\alpha-1)}{\Gamma(\alpha)^2}\left\{(2-\alpha)\delta^2
\Delta^{2\alpha-3}\int_0^{\frac{t_a}{\Delta}}q^{\alpha-1}(1+q)^{\alpha-3}dq+
\Delta^{2\alpha-2}\delta\int_{\frac{t_a}{\Delta}}^{\frac{t_a+\delta}{\Delta}}
q^{\alpha-1}(1+q)^{\alpha-2}dq\right\}.
\label{d4}
\end{eqnarray}
Then, for weak aging $t_a\ll\Delta$, we have
\begin{eqnarray}
\label{d5}
\nonumber
C(t_a,\Delta)&\sim&\frac{2K_\alpha(\alpha-1)}{\Gamma(\alpha)^2}\left\{(2-\alpha)
\delta^2\Delta^{2\alpha-3}\int_0^{\frac{t_a}{\Delta}}q^{\alpha-1}dq+\Delta^{2
\alpha-2}\delta\left[\int_0^{\frac{t_a+\delta}{\Delta}}q^{\alpha-1}dq-\int_0^{
\frac{t_a}{\Delta}}q^{\alpha-1}dq\right]\right\}\\
&=&\frac{2K_\alpha(\alpha-1)}{\alpha\Gamma(\alpha)^2}\left\{(2-\alpha)\delta^2
t^\alpha_a\Delta^{\alpha-3}+\delta\Delta^{\alpha-2}\left[\left(t_a+\delta\right)
^\alpha-t^\alpha_a\right]\right\}.
\end{eqnarray}
In particular, when $t_a=0$ the ACF reads 
\begin{equation}
\label{d6}
C(0,\Delta)\sim\frac{2K_\alpha(\alpha-1)}{\alpha\Gamma{(\alpha)}^2}\delta^{
\alpha+1}\Delta^{\alpha-2},
\end{equation} 
and when $t_a\gg\delta$, we have 
\begin{equation}
\label{d7}
C(t_a,\Delta)\sim\frac{2K_\alpha(\alpha-1)}{\Gamma(\alpha)^2}\delta^2t^{\alpha
-1}_a\Delta^{\alpha-2}.
\end{equation}

Next we consider the case of strong aging ($t_a\gg\Delta$) for different
$\alpha$.

\subsection{Case $1/2<\alpha<3/2$}

For $1/2<\alpha<3/2$, for strong aging ($t_a\gg\Delta$) we use the following
approximation for the first integral in Eq.~(\ref{d4}),
\begin{equation}
\label{d8}
\int_0^{\frac{t_a}{\Delta}}q^{\alpha-1}(1+q)^{\alpha-3}dq\approx\int_0^{\infty}
q^{\alpha-1}(1+q)^{\alpha-3}dq=\mathbb{B}(\alpha,3-2\alpha),
\end{equation}
where the (complete) Beta function is defined by \cite{abramowitz1964}
\begin{equation}
\mathbb{B}(a,b)=\int_0^{\infty}s^{a-1}(1+s)^{-a-b}ds=\frac{\Gamma(a)
\Gamma(b)}{\Gamma(a+b)}.
\end{equation}
The second integral in Eq.~(\ref{d4}) reads
\begin{equation}
\label{d9}
\int_{\frac{t_a}{\Delta}}^{\frac{t_a+\delta}{\Delta}}q^{\alpha-1}(1+q)^{\alpha-2}
dq\sim\int_{\frac{t_a}{\Delta}}^{\frac{t_a+\delta}{\Delta}} q^{2\alpha-3}dq\sim
\left(\frac{\Delta}{t_a}\right)^{3-2\alpha}\frac{\delta}{\Delta}.
\end{equation}
Then, substituting Eqs.~(\ref{d8}) and (\ref{d9}) into Eq.~(\ref{d4}), we have
\begin{equation}
\label{d10}
C(t_a,\Delta)\sim\frac{2K_\alpha(\alpha-1)(2\alpha-1)}{\Gamma(2\alpha)|\cos(\pi
\alpha)|}\delta^2\Delta^{2\alpha-3}.
\end{equation}

\subsection{Case $\alpha>3/2$}

For $\alpha>3/2$ and strong aging ($t_a\gg\Delta$), the first integral in
Eq.~(\ref{d4}) diverges as $t$ tends to infinity. We rewrite the first integral
in two parts,
\begin{equation}
\label{d11}
\int_0^{\frac{t_a}{\Delta}}q^{\alpha-1}(1+q)^{\alpha-3}dq=\int_0^Mq^{\alpha-1}
(1+q)^{\alpha-3}dq+\int_M^{\frac{t_a}{\Delta}}q^{\alpha-1}(1+q)^{\alpha-3}dq,
\quad\mbox{for }M\gg1.
\end{equation}
Then the integral in Eq.~(\ref{d11}) is dominated by the second term and can be
approximated as
\begin{equation}
\label{d12}
\int_0^{\frac{t_a}{\Delta}}q^{\alpha-1}(1+q)^{\alpha-3}dq \sim\int_M^{\frac{
t_a}{\Delta}}q^{\alpha-1}(1+q)^{\alpha-3}dq\sim\int_M^{\frac{t_a}{\Delta}}
q^{2\alpha-4}dq\sim\frac{1}{2\alpha-3}\left(\frac{t_a}{\Delta}\right)^{2\alpha-3}. 
\end{equation}
The second term in Eq.~(\ref{d4}) can be approximated by Eq.~(\ref{d9}). Then,
substituting Eqs.~(\ref{d12}) and (\ref{d9}) into Eq.~(\ref{d4}), we obtain
\begin{equation}
C(t_a,\Delta)\sim\frac{2K_\alpha(\alpha-1)^2}{(2\alpha-3)\Gamma(\alpha)^2}
\delta^2t^{2\alpha-3}_a.
\end{equation}

\subsection{Case $\alpha=3/2$}

For $\alpha=3/2$ the first integral in Eq.~(\ref{d4}) can be solved explicitly,
and for strong aging $t_a\gg\Delta$ we find an approximation with the leading
term
\begin{equation}
\label{d14}
\int_0^{t_a/\Delta}q^{1/2}(q+1)^{-3/2}dq=\ln\left(\frac{t_a}{\Delta}\right)+\ln
\left(1+\frac{\Delta}{t_a}\right)-2\sqrt{\frac{t_a}{t_a+\Delta}}\sim\ln{\left(\frac{t_a}{\Delta}\right)}.
\end{equation}
The second term in Eq.~(\ref{d4}) can be approximated by Eq.~(\ref{d9}) with
$\alpha=3/2$,
\begin{equation}
\label{d15}
\int_{t_a/\Delta}^{(t_a+\delta)/\Delta}q^{1/2}(1+q)^{-1/2}dq\sim\frac{\delta}{
\Delta}.
\end{equation}
Then, substituting Eqs.~(\ref{d14}) and (\ref{d15}) into Eq.~(\ref{d4}), we get
\begin{equation}
\label{d16}
C(t_a,\Delta)\sim\frac{2K_{3/2}}{\pi}\delta^2\ln{\left(\frac{t_a}{\Delta}\right)}.
\end{equation}

To summarize, for strong aging $t_a\gg\Delta$ the ACF of RL-FBM increments for
all $\alpha$ is given by
\begin{equation}
\label{vacf-strong}
C(t_a,\Delta)\sim\left\{\begin{array}{ll}\displaystyle\frac{2K_\alpha(\alpha-1)
(2\alpha-1)}{\Gamma(2\alpha)|\cos(\pi\alpha)|}\delta^2\Delta^{2\alpha-3},\quad &
1/2<\alpha<3/2\\[0.32cm]
\displaystyle\frac{2K_{3/2}}{\pi}\delta^{2}\ln{\left(\frac{t_a}{\Delta}\right)},&
\alpha=3/2\\[0.32cm]
\displaystyle\frac{2K_\alpha(\alpha-1)^2}{(2\alpha-3)\Gamma(\alpha)^2}\delta^2
t_a^{2\alpha-3},&\alpha>3/2\end{array}\right..
\end{equation}
We note that in the regime $1/2<\alpha<3/2$ the ACF becomes stationary and solely
depends on the lag time $\Delta$. The consistency of the simulated and analytical
ACF for RL-FBM increments at strong aging time $t_a\gg\Delta$ in
Eq.~(\ref{vacf-strong}) are shown in Fig.~\ref{Figure5}. A small discrepancy is
observed for the case $\alpha=1.2$ around $\Delta=0.1$ because the relation
$\delta\gg\delta$ is not perfectly fulfilled.

\begin{figure}
\includegraphics[width=0.43\textwidth]{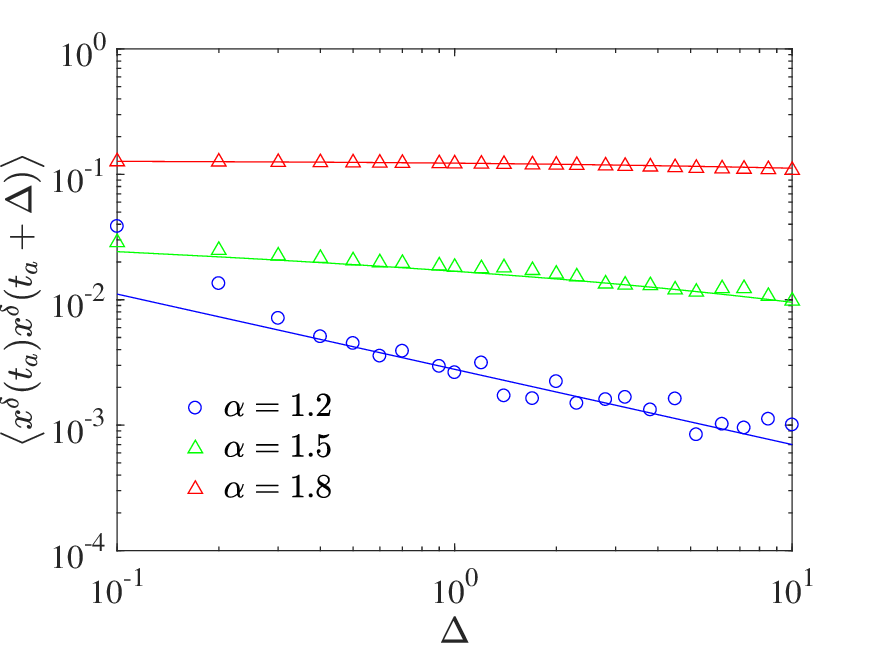}
\setcounter{figure}{0}
\renewcommand{\thefigure}{C\arabic{figure}}
\caption{Simulations (symbols) of the ACF of the increments of RL-FBM at strong
aging $t_a\gg\Delta$. The theoretical results (\ref{vacf-strong}) are represented
by solid lines. Here we set $t_a=20$, $\delta=dt=0.1$, and $K_\alpha=0.5$.}
\label{Figure5}
\end{figure}

\end{document}